\documentclass[10pt,a4paper]{article}
\usepackage{pgf}
\usepackage{tikz}
\usetikzlibrary{arrows,automata}
\usetikzlibrary{arrows,decorations.pathmorphing,backgrounds,positioning,fit,petri}
\usetikzlibrary{positioning,shapes,arrows,arrows.meta}
\tikzstyle{block} = [rectangle, node distance=1.0cm, minimum width=0.5cm, minimum height=0.3cm, draw, thick, 
text width=9em, text centered, rounded corners, minimum height=1.5em]
\tikzstyle{blockg} = [rectangle, minimum width=0.5cm, minimum height=0.8cm, draw, 
text width=5em, text centered, rounded corners, minimum height=1.5em]
\tikzstyle{blockh} = [rectangle, minimum width=0.5cm, minimum height=0.8cm, draw, 
text width=4em, text centered, rounded corners, minimum height=1.5em, very thick,]
\tikzstyle{line} = [draw, -latex',-{Stealth[scale=1.0]},rounded corners,thick]
\tikzstyle{arrow} = [-{Stealth[scale=1.0]},rounded corners,thick]
\usepackage{amsmath,amssymb}
\usepackage{xcolor}
\usepackage{mathrsfs,amsfonts,bm}
\usepackage{hyperref}
\usepackage{graphicx}
\usepackage{enumerate}
\usepackage[algo2e]{algorithm2e}
\usepackage{booktabs}
\usepackage{multirow}

\newcommand{\N}{\mathbb{N}}

\newcommand{\Ne}[1]{\mathbf{#1}}
\newcommand{\ca}[1]{\mathcal{#1}}

\newcommand{\eg}{e.g., }

\newcommand{\ie}{i.e., }
\newcommand{\RU}{\mathbb{R}}
\newcommand{\tn}[1]{\textnormal{#1}}

\newtheorem{rmk}{Remark}
\DeclareMathOperator*{\argmax}{arg\,max}

\usepackage{todonotes}
       
\newcommand{\comment}[1]{\vspace{5 mm}\par \noindent
  \marginpar{\textsc{Comment}} \framebox{\begin{minipage}[c]{0.95 
       \textwidth} \tt #1 \end{minipage}}\vspace{5 mm}\par}

\renewcommand{\comment}[1]{}

\begin{document}

\title{A structure-preserving numerical approach for simulating algae blooms 
	in marine water bodies of western Patagonia\footnote{Article submitted to Ecological Modelling}}

\author{
Pablo~Mata~Almonacid and Carolina~Medel\\[0.4cm]
\small Centro de Investigaci\'on en Ecosistemas de la Patagonia (CIEP)\\
\small Camino Baguales s/n, Coyhaique, 5951601, Chile\\
\small email: \textit{pmata@ciep.cl},~\textit{carolina.medel@ciep.cl}\\[0.2cm]
}

\date{July, 2020}


%
\maketitle

\begin{abstract}
Patagonian fjords' area is one of the largest and less studied estuarine regions in the world, located in the southern coast of Chile. Every one of its water bodies displays a unique hydrodynamic behavior which is strongly determined by the local environmental conditions and has enormous effects on the biogeochemical characteristics of the marine ecosystems. In this context, algal blooms are ecological phenomena of major relevance. They produce strong impacts on both ecosystem's services as well as human activities. Numerical simulation has proved to be a promising tool to understand and anticipate their impacts. Unfortunately at the present, it has not been used for studying algal blooms in this zone. This article focuses on contributing to fill that gap in knowledge by means of proposing a novel numerical model for simulating brief but otherwise intense algal blooms occurring in semi-enclosed marine water bodies of western Patagonia. The proposed model presents a trade-off between complexity and applicability since field-data sparsity in the zone discourages constructing more sophisticated approaches. The model is based on a two-layer description of the water column. The first layer represents the euphotic zone where an embedded biogeochemical model of NPZD-type is used to model a mass-conserving trophic web. High intensity wind drives the water column mixing, introducing an upward flux of nutrients that boosts high rates of primary production in the euphotic zone. A time-dependent Gaussian pulse is used to describe this process. Mass losses due to detritus sinking are also included. Then, the ecosystem's dynamics is represented by means of an externally forced, non-autonomous system of ordinary differential equations which is characterized by strictly positive trajectories but that it is not longer mass-conserving. A structure-preserving time integrator based on a splitting-composition technique is specially designed for solving the system's equations. It is cast as a three-steps algorithm and provides an exact estimations of biomass fluxes. In the first step, a modified Patankar-Runge-Kutta scheme \cite{BurchardEtAl2} is used to solve the unforced NPZD system. Second and third steps consider the effects due to nutrient's pulse and detritus' sinking. Additionally, a genetic algorithm-based tool is used to calibrate the model's parameters in realistic scenarios. Finally, the proposed model is applied to carry out a detailed study of an unusual winter bloom of dinoflagellates in an austral fjord \cite{MonteroEtAl2}. To the best of the authors' knowledge, this case study constitutes the first attempt to model oceanic biogeochemical processes in this world's region.
\end{abstract}

%
\tableofcontents
\newpage
%
%
%
%
\section{Introduction}\label{S.1}
The \textit{Patagonian fjords area} (PFA) is formed by a complex network of interconnected semi-enclosed water bodies which includes fjords, channels, bays and gulfs, located in the southern coast of Chile. It extends from 41.5$^\circ$S (Reloncav\'i Fjord) to 56$^\circ$S (Cape Horn) and constitutes one of the largest estuarine regions in the world \cite{IriarteEtAl1} with more than 3.300 islands, covering an approximate area of 240.000 km$^2$ and 84.000 km of coastline \cite{Pickard1,SilvaVargas1}. FPA's fjords and channels may be described as glacially over-deepened marine basins \cite{HoweEtAl1}. They are partially confined by the local topography which often presents one or more entrance sills separating their waters from the sea. Moreover, fjords and channels frequently receive fresh water discharges coming from local rivers, surface runoff, and ground water flows fed by high rainfall \cite{NeshybaFonseca1} and glaciers \cite{Moffat1,PantojaEtAl1}. Field works indicate that freshwater fluxes from rivers \cite{Romero1} along with the high rainfalls are responsible for the presence of an estuarine circulation \cite{Stigebrandt1} which generates a characteristic stratification of the water column as described in \cite{SchneiderEtAl1}, where the following two basic layers are identified: i) A lower-salinity ($<11-31$ PSU) upper layer of varying depth (5-10 m) \cite{Sievers1}. The seasonal variation of temperature in this layer is strongly influenced by the local conditions of solar irradiance \cite{DavilaEtAl1} and, ii) a lower layer of nutrient-rich waters presenting a more uniform salinity and higher density \cite{SilvaCalvete1}. 

A salient feature of this two-layer structure is the presence of a pronounced vertical salinity gradient between both layers \cite{PerezEtAl1}. The water flows offshore in the top layer and inshore at the bottom layer, contributing to the formation of a transitional marine area \cite{IriarteEtAl1}. Over the continental shelf, semi-diurnal ocean tide is modified due to presence of shallower waters and a higher contribution of nonlinear effects. In spite of their common characteristics, every water body displays a unique behavior in terms of the spatial distribution and temporal variation of currents, salinity and temperature \cite{CastilloValenzuela1,RossEtAl1}. This behavior is determined by the local environmental conditions (e.g., bathymetry, freshwater sources, tidal regime, solar irradiance, etc.) and has enormous effects on the biogeochemical component of the ecosystem dynamics \cite{CastroEtAl1}. Nowadays, the Chilean \textit{National Oceanographic Committee} develops a permanent research program for the multidisciplinary study of PFA's water bodies \cite{SilvaPalma1}. Although collected data have provided valuable scientific information for more than two decades, a knowledge-gap about the ecosystem dynamics of fjords and channels remains being a major obstacle for the implementation of an ecosystem-based management of the zone \cite{IriarteEtAl1,ArheimerNilssonLindstrom}.

PFA also constitutes a crucial source of ecosystem services. In particular, a world class industry of aquaculture has developed in some areas of the northern PFA becoming the key promoter of the social and economical development of the zone \cite{OuteiroVillasante1} as well as the most important consumer of ecosystem services. Nowadays, ensuring the environmental sustainability of aquaculture has become a worldwide issue \cite{BartonFloysand1} since oftenly the industry-ecosystem-community interactions are poorly known, paving the way for public policies that underestimate the long term effects of productive activities on the ecosystem \cite{HosonoEtAl1,BuschmannEtAl2}. 
Therefore, research oriented to understand the coupled physical-biogeochemical dynamics of marine ecosystem becomes a key element to develop sustainable policies for PFA.

From the point of view of the biogeochemical cycles, FPA remains being one of the less studied zones of the world \cite{StrubEtAl1,EscribanoEtAl1}. It presents highly complex marine-terrestrial-atmospheric interactions that may result in high biological production \cite{IriarteEtAl2} involving the exchange of large amounts of matter and energy between terrestrial and maritime ecosystems \cite{CastilloEtAl2}. Estuarine ecosystem dynamics is also strongly influenced by climate change and human activities; se \eg \cite{IriarteEtAl3} and also \cite{TorresEtAl2,LafonEtAl1}. In this context some of the natural phenomena of major relevance in the dynamics of estuarine ecosystem are the algal blooms. They correspond to the uncontrolled proliferation of certain species of phytoplankton as result of complex interactions among biogeochemical, physical and sedimentary processes occurring at different time scales \cite{BerdaletEtAl1}. Among other consequences, they threaten the sustainable production of the aquaculture industry in the PFA \cite{DiazEtAl1}.

Mathematical modeling and computer simulation offer a way to synthesise our understanding of the environment. They serve to study ecosystem's changes under human or climate driven perturbations and to anticipate future scenarios \cite{TettPortillaGillibrandInall1,JickellsEtAl1}. Numerical modeling of marine ecosystems requires to be able to integrate physics, chemistry and biology. On the one hand, specific computer codes for the hydrodynamic component of the ecosystem have been applied to study the waste's dispersion \cite{DoglioliEtAl1}, tidal dynamics \cite{WanEtAl1,PonteCornuelle1}, circulation patterns \cite{HetlandDiMarco1}, wind driven ocean circulation \cite{GhilEtAl1}, flushing time estimation \cite{SadrinasabKampf1}, transport and fate of sediments \cite{DelandmeterEtAl1} and river plumes dynamics \cite{LuShi1,LacroixEtAl1} to name only a few possible applications. On the other hand, numerous efforts in numerical modeling have been focused on the biogeochemical component of the ecosystem's dynamics. The basic goal behind mechanistic models consist in proposing adequate representations of the ecosystem's food web. In general terms, a typical model considers that phytoplankton uses nutrients, light and CO$_2$ to (photo)-synthesize living organic matter while grazing due to zooplankton, mortality, and remineralisation are considered in upper trophic levels. Biogeochemical models have been successfully used as stand alone tools \cite{PortillaTettGillibrandInall1,BurchardEtAl3} or coupled to hydrodynamic modules, see \eg \cite{FennelNeumannBook1,Franks1}. In both cases meteorological, open boundary and nutrient forcing factors need to be defined. Applications include but are not limited to the study of carbon transfer from primary production to mesopelagic fish \cite{Anderson1}, load carrying capacity in bivalve aquaculture \cite{IbarraEtAl1}, ecosystem eutrophication \cite{TettEtAl1}, water quality in estuarine systems \cite{ChenEtAl1}, (harmful) algal blooms \cite{ReigadaEtAl1,BanasEtAl1,ChakrabortyFeudel1}, primary production evaluation \cite{LosEtAl1}, cyanobacteria bloom dynamics \cite{HenseaBurchard1,HenseBeckmann1} to name only a few of them. At present, unfortunately, biogeochemical modeling has not been applied for studying the dynamical behavior of the PFA's ecosystems.

The formulation and computer implementation of mathematical models for marine ecosystems involves a series of challenging tasks among which we highlight: i) it is desirable to be able to reproduce as much as possible the conservation of the dynamical invariants, the so called conserved quantities, \cite[$\S$IV]{HairerEtAlBook1} along with the geometric structure that the system's trajectories may present \cite{HeinleSlawig2,ZhangWang1,EdwardsBrindley1}, and ii) to be able to determine a set of optimal model's parameters for practical case studies \cite{FennelEtAl1}. Regarding to i) recent research efforts have been devoted to the formulation of advanced time stepping methods able to ensure that solution trajectories remain being strictly positive during numerical simulations \cite{BroekhuizenEtAl1,Edwards1} along with enforcing the exact conservation of the total mass of the system \cite{KopeczMeister1,SchippmannBurchard1}. In particular we mention the so called Patankar-Runge-Kutta family of numerical methods which are designed to remain strictly positive and mass-conserving for certain biogeochemical models \cite{BurchardEtAl2,BurchardEtAl4}. We also mention symplectic methods \cite{Marsden2} which, when applied to marine ecosystem models derived from Hamiltonian formalism of mechanics, automatically behave as structure-preserving time integrators \cite{DieleMarangi1}. In what regards to ii), the determination of model's parameters has been identified as a complex task \cite{RuckeltEtAl1,Kriest1} since very oftenly the space where parameters belongs to, may present a complex topology involving multiple sub-optimal regions as it is shown in \cite{CharbonneauKnappReport1}. Additionally, under certain circumstances the model's dynamics may results to be very sensitive to small perturbations in some parameter's values \cite{HeinleSlawig3,OschliesGarcon1,RuckeltEtAl1}. In order to estimate a set of optimal parameters, several methodologies have been developed \cite{HeinleSlawig1}. In particular in ecological modeling of marine ecosystems we recall genetic algorithms \cite{SivanandamDeepaBook1}, Bayesian methods \cite{ArhonditsisEtAl1,ArhonditsisEtAl2} and analytical methods \cite{SchartauEtAl1}.

This article focuses on the formulation of a mathematical model and the development of an appropriate numerical method for the computer simulation of brief but otherwise intense algal blooms occurring in semi-enclosed marine water bodies of western Patagonia. The proposed model presents a trade-off between complexity and applicability since on the one hand the field-data sparsity discourages the implementation of more sophisticated models for marine bodies in the PFA. On the other hand, in spite of its relative simplicity, only a careful design of the time-integration method allows to ensure that discrete approximations to the solution trajectories will retain the geometric structure of their continuous counterparts. Moreover, model's calibration also entails a number of additional difficulties mainly related to a large and topologically complex parameter space with multiple local optima.

The proposed model is based on a two-layer description of a representative volume of the marine environment. The first layer corresponds to the euphotic zone where an embedded biogeochemical model of NPZD-type\footnote{NPZD-model: Corresponds to a four functional-groups biogeochemical model that includes nutrient(N)-phytoplankton(P)-zooplankton(Z)-detritus(D). See \eg \cite{FennelNeumannBook1}.} \cite{FashamEtAl1,HeinleSlawig2,OschliesGarcon1} is used to represent a mass-conserving trophic web. We consider that high intensity winds drive the water column mixing, introducing an upward flux of nutrients that boosts high rates of primary production in the euphotic zone. The nutrient entrainment is described by means of a time-dependent Gaussian pulse. As a result, the ecosystem dynamics corresponds to that of an externally forced, non-autonomous system of ordinary differential equations which is characterized by strictly positive trajectories but that it is no longer mass-conserving. Mass loss is introduced by means of allowing the detritus to sink toward lower water levels. A tailored numerical method is introduced in order to ensure a structure-preserving time integration of the system's equations as well as exact estimations of the biomass' fluxes through the trophic web. The method corresponds to a particularization of the so called splitting-composition techniques and it is cast as a three-steps algorithm. In the first step, a second-order modified Patankar-Runge-Kutta method is applied to the unperturbed NPZD system. The second and third steps consider the effect of the time dependent pulse of nutrients an the biomass loss due to sinking. Both of them are carried out by means of exact integration. A genetic algorithm is suggested as a base-tool for determining an optimal set of model parameters when applied to realistic scenarios. Finally, the proposed model is applied to study an unusual winter bloom of dinoflagellates in an austral fjord (2015) \cite{MonteroEtAl2}. It is worthwhile to mention that to the best of the authors' knowledge, this case study constitutes the first attempt to model coupled physical-biogeochemical processes in the marine ecosystems of this world's region. As result a set of optimal parameters characterizing the biomass fluxes through the food web during an algal bloom is provided. Numerical simulations also allow to study the time scales associated to primary production during the algal bloom.

Article's layout is as follows: $\S$\ref{S.3} focuses on presenting a conceptual framework for the model along with its mathematical formulation. Both the geometric characteristics of the solutions and their conservation properties are analyzed. In $\S$\ref{NumericalMethod} a positively-preserving time integrator able to provide an exact balance of the biomass flux through the trophic web is formulated. The employment of a genetic algorithm is proposed in $\S$\ref{ParameterOptimization} for determining an optimal set for model's parameters. $\S$\ref{NumStudies} is devoted to two case studies. While the first case highlights the algorithm's numerical properties, the second case comprises a detailed description its application to a realistic bloom scenario. Conclusions and further research are presented in $\S$\ref{Conclusions}.           
%
%
%
%
%
\section{Model formulation}\label{S.3}
This section focuses on the conceptual and mathematical formulation of a medium-complexity model able to provide a coarse representation of the most relevant bio-geochemical processes occurring during a marine algal bloom of short duration. The formulation considers the photosynthetical conversion of a pulse of inorganic nutrients into biomass and its subsequent transfer to other trophic levels in a simplified food web. Hydrodynamic processes are driven by infrequent but otherwise intense environmental forcing factors. The proposed model is expected to provide a proper representation for brief scenarios of high primary production occurring in some semi-enclosed water bodies of western Patagonia.     
\subsection{Conceptual model}\label{ConceptualM}
\begin{figure}[h!]
	\begin{center}
		\includegraphics[width=0.9\textwidth]{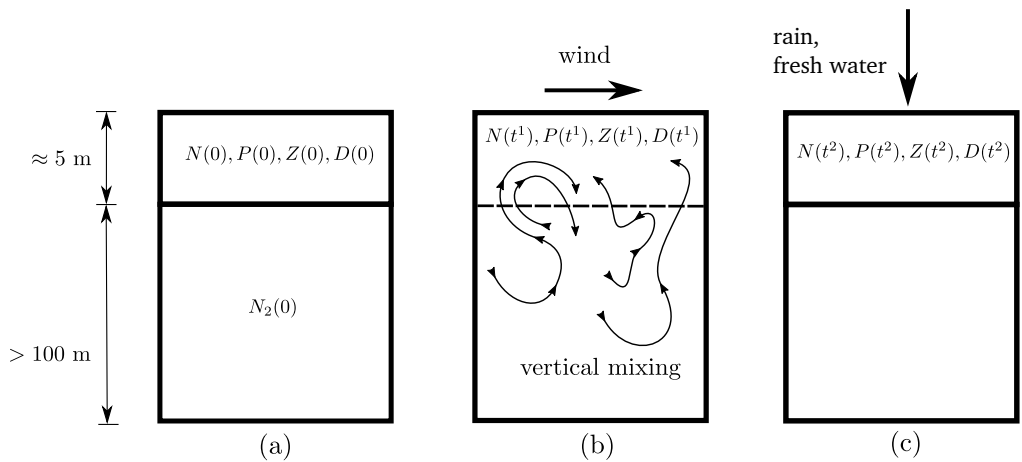}
		\caption{Conceptual model: Two-layer box-model representing the coupled physical and biological dynamics of an algae bloom. While the upper layer (L$_1$) corresponds to the euphotic zone (with \eg 5 m depth), the lower layer (L$_2$) is considerably deeper. $a$) The initial configuration corresponds to a stratified water column with higher nutrient content in L$_2$ ($N_2(0)>N(0)$). Additionally, L$_1$ is characterized by initial values of phytoplankton $P(0)$, zooplankton $Z(0)$ and detritus $D(0)$. $b$) Intense winds trigger vertical mixing in the water column propitiating an upward entrainment of nutrients from L$_2$ into L$_1$. Nutrient enrichment in L$_1$ along with appropriate light and temperature conditions drives the conversion of $N$ to $P$ due to photosynthesis, $P$ to $Z$ due to grazing, $P$ and $Z$ to $D$ due to mortality and $D$ to $N$ due to remineralization. $c$) Later re-stratification occurs due to rains and fresh water discharges. The coupled dynamics of the four functional groups evolves in time in order to reach a long term equilibrium configuration $N(t^2)$-$P(t^2)$-$Z(t^2)$-$D(t^2)$.}
		\label{ConceptualModel}
	\end{center}
\end{figure}

In order to describe the coupled physical-biological dynamics of marine ecosystems experiencing brief but possibly intense algae blooms, we consider a box-model composed of two layers. The first layer represents 1 $\tn{m}^2$ of sea surface with a depth corresponding to that of the euphotic zone at the location of interest. All the biogeochemical processes affecting primary production as well as the short term fluxes of biomass through the food web are assumed to have place in this layer. The second layer is considerably deeper and both, its physical and biological properties are assumed constant for the time scale associated to an algal bloom event. See Figure \ref{ConceptualModel}. A practical computation of the first layer depth based on real data is carried out in section \ref{NumStudies} where the proposed model is applied for simulating a winter bloom of dinoflagellates in an austral fjord.

Regarding to physical properties, the first layer is characterized by a certain initial concentration of nutrients (NO$_3$) measured in [mmol N m$^{-3}$]. This concentration is denoted by $N(0)$ and corresponds to the nutrient content ($N$) at time $t=0$. The influence of temperature and salinity content is currently neglected in the model. The deeper layer is assumed to posses an initially higher concentration of nutrients, $N_2(0)$ thus, yielding to a stratified configuration such as that depicted in Figure \ref{ConceptualModel}-($a$). Frequently, buoyancy forces derived from salinity and temperature dependent stratification tends to compete with turbulent diffusion creating barrier for the upward transfer of nutrients from lower levels into the first layer \cite{BurchardHofmeister1}. Intense fluctuations of environmental factors such as wind, contribute decisively to the loss of the stratification due to water column mixing, introducing nutrient pulses into the euphotic zone and, under appropriate conditions of light availability and temperature, propitiating the oftenly uncontrolled proliferation of phytoplankton ($P$). This phenomenon can be identified with a wind driven marine algal bloom as proposed in \cite{MonteroEtAl2}. See Figure \ref{ConceptualModel}-($b$). Later on, the combined effect of rains and fresh water discharges induces a system's re-stratification as it is shown in Figure \ref{ConceptualModel}-($c$).

From a rough biological perspective, increments in zooplankton ($Z$) biomass are expected to occur due to grazing over $P$, evidencing the flow of organic matter through the ecosystem trophic levels. Moreover, both $P$ and $Z$ experience losses due to natural mortality, contributing to the accumulation of dissolved ($DOM$) and particulate ($POM$) pools of organic matter, which are collectively denoted as detritus ($D$). The bio-geochemical cycle in the first layer is completed by two additional fluxes: i) a nutrient increment due to $Z$ excretions and ii) due to the remineralisation of $D$. Over time scales larger than the bloom duration, $D$ may be subjected to vertical sinking to lower levels in the water column as well.

It is worthwhile to mention that, the following assumptions have been considered: i) a detailed description of the water column hydrodynamics is avoided. Instead an upward pulse of nutrients summarizes the effect of wind driven vertical mixing over the euphotic zone. In this regard, we consider the first layer as a volume that represents the average hydrodynamic and biogeochemical behavior of a considerably larger marine area, ii) an initially stratified water column is considered without making any explicit reference to the depths of the corresponding halocline and themocline curves, iii) blooms are triggered by NO$_3$ increments without taking into account additional dependencies on other factors limiting primary production (\eg limitation by silicate and phosphorus), iv) the functional groups $P$ and $Z$ comprise a huge number of different species belonging to a wide range of morphological scales along with complex predator-prey interactions among them, see \eg \cite{WardDutkiewiczFollows1}. We make an abstraction of such a complexity, reducing those interactions to biomass fluxes among homogeneous functional groups, v) no biological activity is assumed to occur in the second layer.

Some remarkable properties of the model are: 

\begin{enumerate}
	\item[1)] \textit{Mass conservation}: if neither external wind forcing nor detritus sinking are present, the total mass of the system remains invariant in time. On the contrary, if it interacts with the surrounding environment, the mass increment equals the nutrient input minus the biomass loss due to detritus sinking (measured in equivalent units; see $\S$\ref{BiogeochemicalData} for details).
	\item[2)] \textit{Positive system}: we identify the time evolution of the four functional groups with time dependent functions describing the  ecosystem dynamics. Then, it is clear that $N(t)$, $N(t)$, $P(t)$, $Z(t)$ and $D(t)$ must remain greater or equal to zero for any $t\in\RU^+_0$ since negative values are physically unrealistic. Systems with this property are called positive systems (or positively preserving systems) \cite{BurchardEtAl2}.      
\end{enumerate}     
%
%
\subsection{Mathematical model}\label{MathModel}
\begin{figure}[!htbp]	
	\begin{center}
		\begin{tikzpicture}[scale=0.75]
		\node(w) at (-0.7,6) {$\tn I_N(t)$};
		\node(s) at (2,0) { };
		\node[state,circle,very thick] (N) at (2,6) {N};
		\node[state,circle,very thick] (P) at (6,6) {P};
		\node[state,circle,very thick] (Z) at (6,2) {Z};
		\node[state,circle,very thick] (D) at (2,2) {D};
		\draw[line width=.25mm,-{Latex[length=2.5mm]}] (w) -- (N);
		\draw[line width=.25mm,-{Latex[length=2.5mm]}] (D) to node[left] {$\tn I_D(t)$} (s);	
		\draw[line width=.25mm,-{Latex[length=2.5mm]}] (N) to node[above] {\footnotesize $J(N,I)P$} (P);
		\draw[line width=.25mm,-{Latex[length=2.5mm]}] (P) to node[right] {\footnotesize $G(P)Z$} (Z);
		\draw[line width=.25mm,-{Latex[length=2.5mm]}] (Z) to[bend left] node[below,sloped] {$\phi_zZ$} (N);
		\draw[line width=.25mm,-{Latex[length=2.5mm]}] (Z) to node[below] {\footnotesize $(1-\beta)G(P)Z$} (D);    
		\draw[line width=.25mm,-{Latex[length=2.5mm]}] (D) to node[left] {$\gamma_mD$} (N);
		\draw[line width=.25mm,-{Latex[length=2.5mm]}] (P) to[bend right] node[below,sloped] {$\phi_pP$} (D);
		\end{tikzpicture}
	\end{center}
	\caption{Schematic representation of the biogeochemical processes in the euphotic layer. The four functional groups are represented by four state variables: $N$ corresponds to nutrient content, $P$ to phytoplankton, $Z$ to zooplankton and $D$ to detritus. An arrow between two state variables correspond to a biomass's flux between the corresponding trophic levels. The external flux $\tn I_N(t)$ defines a time dependent increment of nutrients associated to wind driven vertical mixing of the water column. Additionally, $\tn I_D(t)$ corresponds to a detritus loss due to sinking towards lower levels in the water column.}  
	\label{MathematicalModel}
\end{figure}
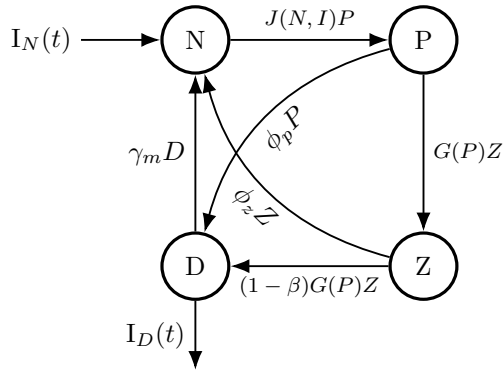

A mathematical model for the two-layer system can be built as an adaptation of the four-components, nutrient-phytoplankton-zooplankton-detritus model ($NPZD$)\cite{OschliesGarcon1,OschliesKoeveGarcon2,HeinleSlawig1,HeinleSlawig2} as it is shown in Figure \ref{MathematicalModel}. The basic idea consists in assuming that under no external forcing the ecosystem dynamics in the euphotic zone is well represented by the solution trajectories of a system of autonomous (\ie time independent) ordinary differential equations (ODE's) describing the conversion of inorganic nutrients into phytoplankton and the subsequent biomass flow through the food web. This process evolves in time until a possible equilibrium state is reached. The upward input of nutrients in the first layer due to wind driven vertical mixing is considered as a time dependent perturbation added to the right hand side (r.h.s) of the balance equation for $N$. Therefore, the new system of ODE's becomes non-autonomous. Finally, biomass transfers toward the lower layer are taken into account by means of an extra flux affecting the r.h.s of the balance equation for $D$. In following detailed explanations are given.       

We assume that the time evolution of the ecosystem is represented by a trajectory of the form
\begin{equation}
\Ne z:[0,T]\rightarrow\RU^{4+},
\end{equation}
where the vector of state variables is given by $\Ne z(t)=(N(t),P(t),Z(t),D(t))$ for all $t$ belonging to a time interval of interest $[0,T]$. In order to keep the model consistent all the state variables are expressed in [mmol N m$^{-3}$]. In that sense, the proposed model can be considered as a Nitrogen-based model. Specific methods to convert $P$, $Z$ and $D$ to the nutrient's unit are given in section \ref{NumStudies}.

\subsubsection{Autonomous NPZD system}\label{AutonomousNPZD}
As explained before, an autonomous, mass conserving and positively preserving $NPZD$ system is obtained by taking $\tn I_N=0$ and $\tn I_D=0$ in Figure \ref{MathematicalModel}. In this case, the system trajectory corresponds to the solution of the following ODE's system 
\begin{subequations}
	\begin{equation}\label{npzd_ode1}
	\dfrac{d\Ne z(t)}{dt}=
	\Ne f_a\big(\Ne z(t);\bm\theta\big),
	\end{equation}
	subjected to initial conditions $\Ne z(0)=(N(0),P(0),Z(0),D(0))$. In the above equation, the four components of the flux vector $\Ne f_a$ are given by  
	\begin{equation}\label{npzd_ode2}
	\tn f_{a_i}(\Ne z;\bm\theta)=
	\sum_{j=1}^{4}
	\left(
	\tn P_{ij}(\Ne z;\bm\theta)-
	\tn D_{ij}(\Ne z;\bm\theta)
	\right),\qquad i=1,\ldots,4.
	\end{equation}
	where $\Ne P(\Ne z;\bm\theta)$ and $\Ne D=\Ne P^t$ are the so called \textit{production} and \textit{destruction} matrices, respectively and $\bm\theta$ is a vector collecting the model parameters as explained below.
	
	\begin{rmk}
		The format chosen to present \eqref{npzd_ode1} and \eqref{npzd_ode2}, even being uncommon, obeys to practical considerations. Production-destruction dynamical systems encompass a wide range of conservative and positive systems arising in several areas of science and in particular to numerous biogeochemical models; see \eg \cite{BurchardEtAl2}. All of them may be expressed in the above format, thus allowing to extend the application of proposed model either to other forms of the flux vector $\Ne f_a$ or to other representations of the trophic web. Moreover, structure-preserving numerical methods for such a systems have receive considerable attention in recent literature \cite{KopeczMeister1} $\blacksquare$ 
	\end{rmk}

	In this work we consider 
	\begin{equation}\label{P_matrix}
	\Ne P(\Ne z;\bm\theta)=
	\begin{bmatrix}
	0 & 0 & \phi_z Z & \gamma_mD\\
	J(N,I)P & 0 & 0 & 0\\
	0 & G(\epsilon,g,P)Z & 0 & 0\\
	0 & \phi_pP & (1-\beta)G(\epsilon,g,P)Z+\phi_z^\ast Z^2 & 0
	\end{bmatrix},
	\end{equation}
	where the \textit{growth rate of phytoplankton} is given by 
	\[
	J(N,I)=\mu_m\frac{N}{k_N+N}\frac{I}{k_I+I}\frac{}{},
	\]
	and depends on both the nutrient content $N$ and on the light availability $I$ which is further explained in $\S$\ref{AlgalBloom} below. Additionally, $\mu_m$ is a growth factor. Clearly, for a large but otherwise fixed valued of $I$ we have that $I(k_I+I)^{-1}\approx 1$ and therefore, $\mu_m$ imposes an upper bound for the rate of conversion of nutrients into phytoplankton since $\lim_{N\rightarrow\infty}J(N,\infty)=\mu_m$. On the other hand, during night it is possible to assume that $I\approx 0$ and thus for any $N>0$, we have that $\lim_{I\rightarrow 0}J(N,I)=0$ providing a corresponding lower bound.

	\begin{table}	
		\begin{centering}
			\begin{tabular}{{l}{c}{c}{l}}
				\hline
				Parameter       & Range   & Units                                    & Definition \\
				\hline
				$k_N$           & $\RU^+$ & $\tn{mmol N m}^{-3}$                     & Half saturation factor for nutrient uptake.\\ 
				$k_I$           & $\RU^+$ & $\tn{W m}^{-3}$                          & Half saturation factor for light absorption.\\	
				$\mu_m$         & $\RU^+$ & $\tn{d}^{-1}$                            & Phytoplankton growth rate factor.\\    
				$\phi_z$        & $\RU^+$ & $\tn{d}^{-1}$                            & Zooplankton linear loss rate.\\
				$\phi_z^\ast$   & $\RU^+$ & $\tn{m}^3(\tn{mmol N d})^{-1}$           & Zooplankton quadratic loss rate.\\	
				$\phi_p$        & $\RU^+$ & $\tn{d}^{-1}$                            & Phytoplankton linear loss rate.\\
				$\gamma_m$      & $\RU^+$ & $\tn{d}^{-1}$                            & Remineralization rate for $D$.\\
				$\beta$         & [0,1]   & --                                       & Assimilation efficiency of zooplankton.\\
				$\epsilon$      & $\RU^+$ &	$\tn{m}^6(\tn{mmol N})^{-2}\tn{d}^{-1}$	 & Grazing encounter rate.\\
				$g$             & $\RU^+$ & $\tn{d}^{-1}$                            & Maximum grazing rate.\\
				$\kappa$        & $\RU^+$ & $\tn{d}^{-1}$                            & loss rate for $D$.\\
				$a,b,c$         & $\RU^+$ & --                                       & Gaussian function parameters.\\
				\hline
			\end{tabular}
			\caption{Model's parameters including their definitions, units and ranges of admissible values.}
			\label{TableParameters}
		\end{centering}
	\end{table}
	
	Moreover, in \eqref{P_matrix} a Holling type III function \cite{FashamEtAl1} is used for the \textit{zooplankton grazing ratio} function $G(\epsilon,g,P)$, which given by
	\begin{equation}
	G(\epsilon,g,P)=\frac{g\epsilon P^2}{g+\epsilon P^2},
	\end{equation}
	where $\epsilon$ and $g$ are parameters such that $\lim_{P\rightarrow\infty}G(\epsilon,g,P)=g$ provides an upper bound for the grazing ratio. The remaining model's parameters appearing in \eqref{P_matrix} are summarized in Table \ref{TableParameters} where their definitions along with their units of measure and range of admissible values are specified. For sake of simplicity, we collect the model's parameters in a single vector    
	\begin{equation}\label{Parameters1}
	\bm\theta:=(k_N,k_I,\mu_m,\phi_z,\phi_z^\ast,\phi_p,\gamma_m,\beta,\epsilon,g,\kappa).
	\end{equation}
	Model's behavior is strongly determined by the value of $\bm\theta$ \cite{HeinleSlawig3}. Section \ref{ParameterOptimization} is devoted to model calibration and parameter optimization. 
	
	In what regards to the geometric structure of solutions, we have that the mathematical model inherits the conservation properties of the conceptual model. In particular, we have that
	\begin{enumerate}
		\item[i)] The total mass is exactly conserved. It can verified from the fact that
		\begin{equation}\label{MassConservation}
		0=\frac{d\Ne z}{dt}\cdot\Ne 1=
		\Ne f_a\cdot\Ne 1=
		\frac{dN}{dt}+\frac{dP}{dt}+\frac{dZ}{dt}+\frac{dD}{dt},  
		\end{equation}
		where $\Ne 1=(1,1,1,1)$ and equations \eqref{npzd_ode2} and \eqref{P_matrix} have been taken into account. This property implies that the time evolution of the state variables remains confined to a linear manifold defined by the constraint equation $\Ne z(t)=\Ne z(0)$ for all $t\in[0,T]$.
		\item[ii)] The solution trajectories are positive. This property means that if the system \eqref{npzd_ode1} is subjected to positive initial conditions $\Ne z(0)$ then the time evolution of the state variables $\Ne z(t)$ remains greater or equal to zero for each $\tn z_i$, $i=1,\ldots,4$ and for all $t\in[0,T]$. A proof of this property for any dynamical systems in production-destruction form can be found in \cite[$\S$1]{BurchardEtAl2}.  
	\end{enumerate}
	The numerical method proposed in $\S$\ref{NumericalMethod} is designed to exactly preserve these properties.
\end{subequations}

%
%
\subsubsection{Nutrient input due to vertical mixing}
\begin{subequations}
	
	An upward input of nutrients in the upper layer due to a wind driven vertical mixing is considered by means of modifying the balance equation for $N$ according to 
	\begin{eqnarray}
	\frac{dN}{dt}
	&=&
	\tn f_{a_1}(\Ne z;\bm\theta)+\tn{I}_N(t)
	\nonumber
	\\
	&=&
	-J(N,I)P+\phi_zZ+\gamma_mD+\tn{I}_N(t),
	\label{dNdt_I_N}
	\end{eqnarray}
	where the nutrients' pulse $\tn{I}_N(t)$ is described by a time-dependent Gaussian function
	\begin{equation}\label{GaussianF1}
	\tn{I}_N(t)=a \exp\left(-\frac{(t-b)^2}{2c^2}\right),
	\end{equation}
	with $a,b,c$ being positive scalars controlling the pulse's amplitude, position and width, respectively. Specific values for $a$, $b$ and $c$ are problem dependent and some guidelines for their estimation are given in $\S$\ref{AlgalBloom}.  
\end{subequations}
%
%
\subsubsection{Detritus loss due to vertical sinking}
\begin{subequations}
	Following \cite{HeinleSlawig1}, we consider the precipitation of $D$ from the euphotic zone to deeper levels in the water column. This process is particularly important if the time interval of interest is not restricted to the duration of an algal bloom and a long term characterization of the ecosystem dynamics is required. Then, an extra term is added to the r.h.s of the balance equation for $D$, namely
	\begin{eqnarray}
	\frac{dD}{dt}
	&=& 
	\tn f_{a_4}(\Ne z;\bm\theta)+\tn{I}_D(\Ne z;\bm\theta)
	\nonumber
	\\
	&=&
	\phi_pP+(1-\beta)G(\epsilon,g,P)Z+\phi_z^\ast Z^2-\gamma_mD+\tn{I}_D(\Ne z;\bm\theta)
	\label{Loss1}
	\end{eqnarray}
	where 
	\begin{equation}\label{Loss2}
	\tn{I}_D\left(\Ne z(t);\bm\theta\right)=
	\left\{
	\begin{array}{ll}
	-\kappa\left(D(t)-D^\ast\right),&\tn{if}\quad D(t)\geq D^\ast\\
	0, & \tn{otherwise}
	\end{array}
	\right.,
	\end{equation}
	is an extra flux due to detritus sinking. In the above equation $D^\ast$ provides a lower bound for detritus concentration and $\kappa>0$ is an extra parameter which can be identified with an exponential decay constant. Appropriate values for $D^\ast$ are problem dependent and an example is discussed in $\S$\ref{AlgalBloom}. 
	\begin{rmk} It is worthwhile noting that the externally forced $NPZD$-model shown in Figure \ref{MathematicalModel} allows to accommodate some other external interactions with the environment such as for example zooplankton depredation by means of adding an extra flux $\tn I_Z(t,\Ne z;\bm\theta)$ on the right of functional group $Z$ $\blacksquare$
	\end{rmk}
\end{subequations}
%
%
\subsubsection{Conservation properties}\label{ConservationPContinuum}
\begin{subequations}
	Finally, the dynamical behavior of the two-layer model under the action of time dependent external forcing is described by the following non-autonomous system of ODE's
	\begin{equation}\label{npzd_na_ode1}
	\frac{d\Ne z(t)}{dt}=
	\Ne f_a(\Ne z(t);\bm\theta)+
	\Ne f_b(t)+
	\Ne f_c(\Ne z(t);\bm\theta),
	\end{equation}
	subjected to initial conditions $\tn z_i(0)>0$, $i=1,\ldots,4$. 
	
	In the above equation $\Ne f_a(\Ne z(t);\bm\theta)$ was given in \eqref{npzd_ode2} and \eqref{P_matrix},
	\begin{equation}
	\Ne f_b(t)=\big(\tn{I}_N(t),0,0,0\big)
	\end{equation}
	is a non-autonomous flux vector representing a Gaussian pulse of nutrients into the first layer due to a wind driven mixing of the water column and
	\begin{equation}
	\Ne f_c(\Ne z(t);\bm\theta)=\big(0,0,0,\tn{I}_D(\Ne z(t);\bm\theta)\big)
	\end{equation}
	takes into account the biomass loss due to the detritus precipitation.

	Repeating \eqref{MassConservation} for \eqref{npzd_na_ode1} it is possible to see that
	\begin{eqnarray}
	\frac{d\Ne z}{dt}\cdot\Ne 1
	&=&
	\Ne f_a\cdot\Ne 1+\Ne f_b\cdot\Ne 1+\Ne f_c\cdot\Ne 1
	\nonumber
	\\
	&=&
	\tn{I}_N(t)+\tn{I}_D(\Ne z(t);\bm\theta)
	\nonumber
	\\
	&=&
	\left\{
	\begin{array}{ll}
	\tn{I}_N(t),                               & \tn{if}\quad D(t)<D^\ast\\
	\tn{I}_N(t)-\kappa\left(D(t)-D^\ast\right) & \tn{otherwise}.
	\end{array}
	\right.,
	\label{MassBalance}   
	\end{eqnarray}
	which shows that total mass is no longer an invariant of the dynamics. 
	
	Regarding to the positive character of the solutions and assuming tha $D^\ast>0$, two situations need to be considered: i) if $D(t)<D^\ast$ we have that $N(t)>0$ for all $t$ since \ref{GaussianF1} shows that $\tn{I}_N(t)>0$ and \eqref{dNdt_I_N} holds, ii) if $D(t)>D^\ast$ \eqref{Loss2} shows that an exponential decay of detritus due to precipitation occurs until $0<D(t)=D^\ast$. Therefore, both $\tn I_N$ and $\tn I_D$ preserve the positive character of the solutions.  

\end{subequations}

%
%
%
%
\section{Numerical method}\label{NumericalMethod}
\begin{subequations}
	As it has been highlighted in $\S$\ref{S.1} the formulation of the so called structure-preserving methods able to conserve as much as possible of the geometric properties that solutions of ODE systems may present, have gained considerable attention in computational biogeochemistry because their superior performance in numerical simulations as well as their stability properties \cite{BurchardEtAl2,KopeczMeister1}. Section \ref{S.3} shows that in our case, structure-preservation reduces to ensuring positive solution trajectories along with the exact fulfillment of the mass balance \eqref{MassBalance}. This section is devoted to the construction of a structure-preserving numerical method for the two-layer model.

	For constructing a numerical approximation to the solutions of \eqref{npzd_na_ode1} we apply an \textit{operator splitting method} \cite{HairerEtAlBook1}. To this end we start by considering a conforming partition of the time interval $[0,T]$ which is characterized by an integer $N\geqslant 1$ and a sequence $\{t^n\}_{n=0,\ldots,N}$ of real numbers such that $0=t^0<\cdots<t^n<\cdots<t^N=T$ and where $h=t^{n+1}-t^n$ is the constant time step. 
	
	Then, system \eqref{npzd_na_ode1} is split into the following simpler systems
	\begin{equation}\label{Sub-systems}
	\frac{d\Ne z(t)}{dt}=
	\Ne f_a\big(\Ne z(t);\bm\theta\big),
	\qquad
	\frac{d\Ne z(t)}{dt}=
	\Ne f_b(t),
	\qquad
	\frac{d\Ne z(t)}{dt}=
	\Ne f_c\big(\Ne z(t);\bm\theta\big), 
	\end{equation}
	subjected to appropriate initial conditions.

	The first system corresponds to an autonomous mass-conserving system of NPZD-type; see \eqref{npzd_ode1} and \eqref{npzd_ode2}. The second one takes into account a time-dependent flux of nutrients due to vertical mixing and the third system considers biomass losses due to detritus sinking. Thus, the splitting technique treats separately mass conserving processes from the non-conserving ones.

	Next, we consider the following maps
	\begin{equation}\label{DiscreteMaps}
	\Ne\Phi_h^a,\Ne\Phi_h^b,\Ne\Phi_h^c:\RU^{4+}\rightarrow\RU^{4+},
	\end{equation}
	which provide consistent approximations to the exact flow of the systems \eqref{Sub-systems}. According to \cite{HairerEtAlBook1}, given an initial condition $\Ne z(0)$ at time $t=0$, a first-order approximation to the exact flow of \eqref{npzd_na_ode1} at time $t=h$, $\Ne z(h)$, can be constructed as
	\begin{equation}\label{Numerical_flux_1}
	\Ne\Phi_h\big(\Ne z(0)\big)=
	\big(\Ne\Phi_h^c\circ\Ne\Phi_h^b\circ\Ne\Phi_h^a\big)\big(\Ne z(0)\big) 
	=\Ne z(h)+\ca O\big(h^2\big).
	\end{equation}  
	The above approximation in fact defines an one-step method based on the composition of the mappings given in \eqref{DiscreteMaps}.
	\begin{rmk}
		In some situations such as long term simulations of ecological marine processes, higher order methods are preferable in order to provide a better control on cumulative numerical errors. Such a methods can be formulated following the composition techniques described in \cite{Strang1}. In particular, the mapping
		\[
		\Ne\Phi_h\big(\Ne z(0)\big)=
		\big(\Ne\Phi_{h/2}^a\circ\Ne\Phi_h^b\circ\Ne\Phi_h^c\circ\Ne\Phi_{h/2}^a\big)\big(\Ne z(0)\big), 
		\]
		provides a second-order method for the system \eqref{npzd_na_ode1} $\blacksquare$ 
	\end{rmk}
	In the next sections we provide explicit expressions for the mappings $\Ne\Phi_h^a$, $\Ne\Phi_h^b$ and $\Ne\Phi_h^c$.
\end{subequations}
%
%
\subsection{Autonomous NPZD-system ($\Ne\Phi_h^a$)} 
\begin{subequations}
	A numerical method for the first system in \eqref{Sub-systems} is given by the second-order, unconditionally positive and conservative \textit{modified Patankar-Runge-Kutta} (MPRK) method. This time stepping scheme is a member of a large class of methods originally presented in \cite{BurchardEtAl2} for conservative systems of ODE's in production-destruction form. Posteriorly, the methods were generalized in \cite{KopeczMeister1}. Given $\Ne z^k$ as an approximation to $\Ne z(t^k)$, the MPRK method determines $\Ne z^{k+1}\approx\Ne z(t^{k+1})$ in two steps. The solution procedure requires to solve a linear system of equations thus avoiding iterative methods associated to non-linear systems. 
	
	In the first step the method maps $\Ne z^{k}$ to a fictitious configuration $\Ne z^{p}$ according to   
	\begin{equation}\label{LS1}
	\Ne z^{p} =\Ne\Omega(\Ne z^{k})^{-1}\,\Ne z^n,  
	\end{equation}
	where the entries of the configuration-dependent matrix $\Ne\Omega(\Ne z^{k})$ are given by
	\begin{eqnarray*}
		\Omega_{ii}
		&=&
		1+\frac{h}{\tn z_i^n}
		\sum_{j=1}^4\tn D_{ij}(\Ne z^n),
		\qquad
		i=1,\ldots,4
		\\
		\Omega_{ij}
		&=&
		-\frac{h}{\tn z_j^n}\tn P_{ij}(\Ne z^n),
		\qquad
		i,j=1,\ldots,4,\quad i\ne j.
	\end{eqnarray*}
	
	Finally, in the second step $\Ne z^{k+1}$ is computed as
	\begin{equation}
	\Ne z^{n+1} = \Ne M(\Ne z^{k},\Ne z^{p})^{-1}\,\Ne z^n,  
	\end{equation}
	where the entries of $\Ne M(\Ne z^{k},\Ne z^{p})$ are given by
	\begin{eqnarray*}
		\tn M_{ii}
		&=&
		1+\frac{h}{2\,\tn z_i^p}
		\sum_{j=1}^4
		\left(
		\tn D_{ij}(\Ne z^p)+\tn D_{ij}(\Ne z^n)
		\right),
		\qquad
		i=1,\ldots,4
		\\
		\tn M_{ij}
		&=&
		-\frac{h}{2\,\tn z_j^p}
		\left(
		\tn P_{ij}(\Ne z^p)+\tn P_{ij}(\Ne z^n)
		\right),
		\qquad
		i,j=1,\ldots,4,\quad i\ne j.
	\end{eqnarray*}
	We recall that the unconditional positiveness along with the conservative character of the MPRK method guarantees that for all $h>0$: i) $\Ne z^{k+1}>0$ for all $\Ne z^{k}>0$, and ii) the following condition holds for all $k\in\N$,
	\begin{equation}\label{ExactMassConservation}
	\sum_{i=1}^{4}(\tn z_i^{n+1}-\tn z_i^{n})=0.  
	\end{equation}
	Formal proofs for the previous statements can be found in \cite{KopeczMeister1}.
\end{subequations}
%
%
\subsection{Non-autonomous system ($\Ne\Phi_h^b$)}
\begin{subequations}
	The one-step method $\Ne\Phi_h^b$ maps $\Ne z^k$ to $\Ne z^{k+1}$ in such a way that $\Ne z^{k+1}$ is a consistent approximation to the value $\Ne z(t^{k+1})$ generated by exact flow of the second system of \eqref{Sub-systems} when subjected to the initial condition $\Ne z(t^k)=\Ne z^k$. 
	
	To build the method we note that the solution of the second system of \eqref{Sub-systems} is given by
	\begin{equation}\label{Phi_b}
	N(t^{k+1})=N^k+\int_{t^k}^{t^{k+1}}\hspace{-0.5cm}\tn I_N(s)ds,\quad P(t^{k+1})=P^k,\quad Z(t^{k+1})=Z^k,\quad D(t^{k+1})=D^k,
	\end{equation}
	where the first term admits the exact solution
	\begin{equation}\label{ExactN}
	\tn N(t^{k+1})=N^k+
	ac\sqrt{\frac{\pi}{2}}
	\left(\tn{erf}\left[\frac{t^{k+1}-b}{\sqrt{2}c}\right]-
	\tn{erf}\left[\frac{t^k-b}{\sqrt{2}c}\right]\right),
	\end{equation}
	where $\tn{erf}:\RU\rightarrow[-1,1]$ is the Gauss error function. 
	
	Therefore, it suffices to define $\Ne\Phi_h^b$ as a mapping producing a sequence of discrete values that are coincident with those given by \eqref{Phi_b} and \eqref{ExactN}.
\end{subequations}
%
%
\subsection{Detritus sinking ($\Ne\Phi_h^c$)}
\begin{subequations}
	To construct $\Ne\Phi_h^c$ we proceed analogously to the previous case. The exact solution of the third system of \eqref{Sub-systems} is given by
	\begin{equation}\label{Loss3a}
	N(t^{k+1})=N^k,\quad P(t^{k+1})=P^k,\quad Z(t^{k+1})=Z^k,
	\end{equation}
	along with
	\begin{equation}\label{Loss3b}
	D(t^{k+1})=
	\left\{
	\begin{array}{ll}
	D^\ast+\exp\left(-\kappa h\right)(D^k-D^\ast),&\tn{if}\quad D^k\geq D^\ast\\
	D^k, & \tn{otherwise}
	\end{array}
	\right..
	\end{equation}
	Therefore, we define $\Ne\Phi_h^c$ as the following discrete mapping 
	\[\Ne z^{k+1}=\Ne\Phi_h^c(\Ne z^k)=\Ne z(t^{k+1}),\]
	where $\Ne z(t^{k+1})$ is computed according to \eqref{Loss3a} and \eqref{Loss3b}.   
	
	From the above expression it is clear that in each time step the detritus loss due to vertical sinking can be computed as  $D^k-D^{k+1}\geqslant0$ and that $\lim_{k\rightarrow\infty}D^k=D^\ast$.
\end{subequations}
%
%
\subsection{Geometric properties of $\Ne\Phi_h$}\label{ConservationPDiscrete}
\begin{subequations}
	The geometric properties of the discrete flow generated by $\Ne\Phi_h$	are as follows,
	\begin{enumerate}
		\item[i)] Discrete trajectories are strictly positive. This property can be verified by taking into account that $\Ne\Phi_h$ is obtained according to \eqref{Numerical_flux_1} by composition of $\Ne\Phi_h^a$, $\Ne\Phi_h^b$ and $\Ne\Phi_h^c$. On the one hand, $\Ne\Phi_h^a$ is strictly positive since it corresponds to a MPRK method. Moreover, the composition $(\Ne\Phi_h^b\circ\Ne\Phi_h^a)$ retains this property since $\Ne\Phi_h^b$ only modifies $N^{k+1}$ by adding a positive term. On the other hand, the subsequent composition with $\Ne\Phi_h^c$ is also strictly positive since it only modifies $D^{k+1}$ if it is greater than $D^\ast>0$ forcing it to converge asymptotically to this value.        
		\item[ii)] Mass balance is exactly accounted for. The MPRK method ensure the exact conservation of the total mass as stated in \eqref{ExactMassConservation}. However, the further composition with $\Ne\Phi_h^c$ and $\Ne\Phi_h^c$ modifies this property according to
		\begin{equation}\label{MassBalanceDiscrete}
		\sum_{i=1}^{4}(\tn z_i^{n+1}-\tn z_i^{n})=
		\left\{
		\begin{array}{ll}
		\left(D^k-D^\ast\right)
		\left(\dfrac{1}{e^{\kappa h}}-1\right)
		+f(t^k,t^{k+1}),&\tn{if}\quad D^k\geq D^\ast\\
		f(t^k,t^{k+1}), & \tn{otherwise}
		\end{array}
		\right.,
		\end{equation}
		where $f(t^k,t^{k+1})>0$ corresponds to the second term on the r.h.s of \eqref{ExactN}. 
		
		The above expression provides a discrete version of the ecosystem's biomass balance. The first term of the r.h.s takes into account the mass loss due to detritus sinking and the second term the biomass increment due to nutrient enrichment. This balance results to be exact since the r.h.s of \eqref{MassBalanceDiscrete} is exactly computed in \eqref{Phi_b}, \eqref{ExactN} and \eqref{Loss3b}.
	\end{enumerate}	
\end{subequations}
%
%
%
%
\section{Parameters optimization}\label{ParameterOptimization}
As mentioned in $\S$\ref{AutonomousNPZD}, the model's behavior strongly depends on the values chosen for the set of parameters collected in $\bm\theta$ \eqref{Parameters1}. The ranges of admissible values for the parameters were given in Table \ref{TableParameters}. From this table we note that any admissible $\bm\theta$ must be an element of the parameter space $\ca S:=\RU^{7+}\times[0,1]\times\RU^{3+}$.   

In general, the task of finding a specific $\bm\theta^\ast\in\ca S$ such that the model's output fits experimental data up to certain precision is widely recognized as non-trivial and even very difficult (see \eg \cite{RuckeltEtAl1}) since $\ca S$ may be a really large space possessing a complex topology that involves multiple sub-optimal regions. Sometimes even when an optimal set of parameters may be found, considerable discrepancies with experimental observations might be present, revealing some intrinsic limitations in the model's ability to simulate the physical phenomenon. Additionally, the model's behavior may suffer significant modifications under small perturbations of the most sensible parameters. In this work we make use of the so called \textit{genetic algorithms} (GA) \cite{SivanandamDeepaBook1} for performing the model's parameters calibration. In following we formally pose the optimization problem and then briefly introduce the GA methodology used in $\S$\ref{AlgalBloom} to calibrate the model in an application problem of ecological significance in marine sciences.   
%
%
\subsection{Parameters optimization problem}
\begin{subequations}
	Suppose we are given with a set of observational data $\left\{\Ne z(t^{l})\right\}_{l=1}^L$. For example, $\tn z_1(t^1)$ corresponds to the nutrient content in the euphotic zone measured experimentally at time $t^1$ and the same applies for the other components of $\Ne z(t)$ at time instants $\left\{t^l\right\}_{l=1}^L$.  
	
	The problem of finding an optimal set of model's parameters fitting a given set of observational data is equivalent to  
	\begin{equation}\label{Opt1}
	\argmax_{\bm\theta\in\ca S}\Gamma(\bm\theta)=
	-\sum_{i=1}^{4}w_i
	\left[
	\sum_{l=1}^{n}\left(\tn z_i(t^l)-\tn z_i^l(\bm\theta)\right)^2
	\right],
	\end{equation}
	\ie to find $\bm\theta^\ast\in\ca S$ such that the cost function $\Gamma(\bm\theta)$ has a global maximum in the parameter space. In the above equation $\{\Ne z^l\}_{l=1,\ldots,L}$ corresponds to the numerical output and $\{w_j\}_{j=1}^4$ to a set of positive scalars that takes into account relative weight of each functional group in the construction of  $\Gamma(\bm\theta)$. Moreover, in this work we consider the condition $\sum_{i=1}^4w_i=1$.
\end{subequations}
%
%
\subsection{Genetic algorithm based optimization}\label{ParameterOptimizationGA}
Genetic algorithms for parameter optimization are heuristic search methods that mimic natural selection. In the process, fittest individuals \ie individuals better adapted to their environment, are selected for reproduction in order to produce offspring for the next generation of a breeding population. Evolution also requires the offspring inherits of the parent's superior characteristics (heredity) along with the existence of a fitness spectrum among population members (variability). See \eg \cite{CharbonneauKnappReport1,SivanandamDeepaBook1} for a throughout account on the subject.  

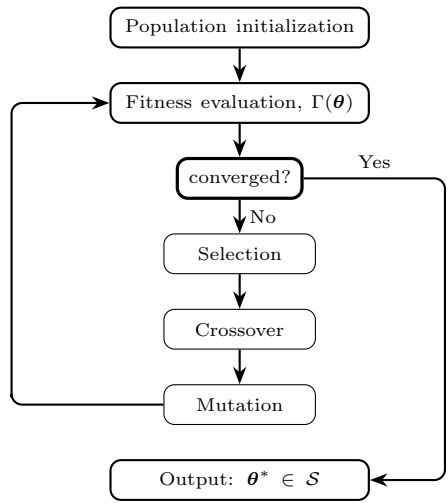
\begin{figure}[!h]	
	\begin{center}
		\begin{tikzpicture}[auto]
		\node [block] (init) {\scriptsize Population initialization};
		\node [block, below of=init] (second) {\scriptsize Fitness evaluation, $\Gamma(\bm\theta)$};
		\node [blockh, below of=second] (conver1) {{\scriptsize converged?}};
		\node [blockg, below of=conver1] (third) {\scriptsize Selection};;
		\node [blockg, below of=third] (fourth) {\scriptsize Crossover };
		\node [blockg, below of=fourth] (fifth) {\scriptsize Mutation};   
		\node [block, below of=fifth] (output) {\scriptsize Output: $\bm\theta^\ast\in\ca S$};
		\coordinate[left of=fifth] (b1);  
		\coordinate[left of=second] (f1);
		\coordinate[right of=conver1] (e1);
		\coordinate[right of=output] (o1);  
		\path [line] (init) -- (second);
		\path [line] (second) -- (conver1);
		\path [line] (conver1) --node {\scriptsize No} (third);
		\path [line] (third) -- (fourth);
		\path [line] (fourth) -- (fifth); 
		\path [line] (fifth) -| ([xshift=-2.0cm]b1) -- ([xshift=-2.0cm]f1)-- (second);
		\path [line] (conver1) -| node [near start] {\scriptsize Yes} ([xshift=1.7cm]o1) -- (output);
		\end{tikzpicture}     
	\end{center}
	\caption{Flowchart of a typical GA for parameters optimization. An initial population of chromosomes is obtained by means of a random sampling of $\ca S$. Every chromosome corresponds to the binary encoding of a parameters vector. Evaluating the optimization function $\Gamma(\bm\theta)$ allows to construct a ranking of chromosomes according to their fitness. If an optimal set of parameters is obtained (convergence checking) the algorithm stops. On the contrary, a selection of individuals for reproduction is carried out based on their mating probability ($p_m$). Pairs of selected chromosomes interchange genes (fragments of binary code) according to a crossover probability ($p_{cross}$) and produce the next generation of breeding population. Chromosomatic mutations have place according to certain probability ($p_{mut}$). Different mutation modes and reproduction plans can be implemented. Iterations are carried out until a certain convergence tolerance is attained. See \eg \cite{CharbonneauKnappReport1}.}  
	\label{GA_Flowchart}
\end{figure}

In practice, the implementation of a GA-based optimization requires to define: i) a \textit{fitness function} for optimization which corresponds to $\Gamma(\bm\theta)$ in \eqref{Opt1}, ii) a \textit{population of chromosomes} that appropriately encode different versions of $\bm\theta$ in \eg binary code. This population can be initialized as a random sampling of $\ca S$, iii) a mechanism to select which chromosomes will reproduce. Basically this process corresponds to build a ranking of chromosomes according to their fitness and to assign to each one of them a reproduction probability according to its ranking's position, iv) a \textit{crossover procedure} to produce a next generation of chromosomes. This operation involves a stochastic procedure for determining the chromosome's segments to be interchanged during mating, v) a procedure to enforce the \textit{random mutation} of some chromosomes in a new generation, vi) a \textit{decoding procedure} to convert the binary data stored in the chromosomes to numerical parameters, and vii) appropriate tolerances for convergence's checking during iterative computations. An overview of a typical flowchart for GA algorithms is sown in Figure \ref{GA_Flowchart}.   

Among the main advantages of GA are: i) they work well for both continuous and discrete optimization functions avoiding the computation of derivatives, ii) they are reasonably efficient and fast, iii) their algorithmic structure allows both the serial and parallel implementations, iv) they provide approximate solutions even in the case of large parameter spaces.  

GA's also suffer from some shortcomings: i) fitness evaluation is frequently required during computations. This may increase ostensibly the computing time in some cases, ii) due to the probabilistic nature of the selection, crossover and mutation processes, the computation of a global optimum is not guarantied, and iii) in certain problems involving topologically intricate parameter spaces, the algorithm might stay iterating around local optima.   
%
%
%
%
%
\section{Numerical studies}\label{NumStudies}
In this section the proposed model is used to conduct two kind of studies. In the first example, the structure-preserving properties of the numerical method are verified along with discussing some of its characteristics. The second example corresponds to a case study. The model is applied for studying an infrequent winter's bloom of dinoflagellates in a semi-enclosed marine water body of western Patagonia. This bloom scenario has been chosen due to it was recently reported in \cite{MonteroEtAl2} as a possible consequence of climate change on the ecosystems' dynamics of austral fjords and channels. To this end, a GA-based calibration of the model is carried out considering observational data in order to characterize the biomass' flow through the food web. Numerical outputs allow making estimations of: i) both primary production and phytoplankton grazing rates, ii) the time scales associated to observed peaks in $N$, $P$ and $Z$ during algae blooms, and iii) the volume of organic matter photo-synthesized in the euphotic zone.
%
%
\subsection{Model's properties}
\begin{subequations}
	As it has been described in $\S$\ref{ConservationPContinuum} and $\S$\ref{ConservationPDiscrete} the proposed time integrator generates positive discrete trajectories along with an exact estimation of the time evolution of the system's biomass. In following the second property is verified in a numerical experiment since the first one is an algebraic property of the algorithm. To this end, we consider a two-layer model with initial conditions specified in the first column of Table \ref{FieldDataConverted}; a first layer's depth equal to 5 [m] and a light availability function as explained in $\S$\ref{solar_irradiance}. Model's parameters are taken from the second row of Tables \ref{OptimalParameters1} and \ref{OptimalParameters2}. Two cases are considered: i) a set of nutrient pulses are specified through the parameters: $b=0.5$, $c=0.424$ and the third parameter belonging to $\left\{a_i\right\}_{i=1}^4=\{15.0, 18.0, 21.0, 24.0\}$ along with a sinking's constant $\kappa=0.0$; see \eqref{GaussianF1}, and ii) a single pulse ($a=21.0$, $b=0.5$ and $c=0.424$) along with a sinking's constant belonging to the set $\left\{\kappa_i\right\}_{i=1}^4=\{0.000,0.025,0.050,0.100\}$. The time evolution of the total biomass per unit volume is shown in Figure \ref{NumericMassBalance}-$(a)$ and $(b)$ for both cases.      
	
	\begin{figure}[!h]
		\begin{center}
			\includegraphics[width=0.85\textwidth]{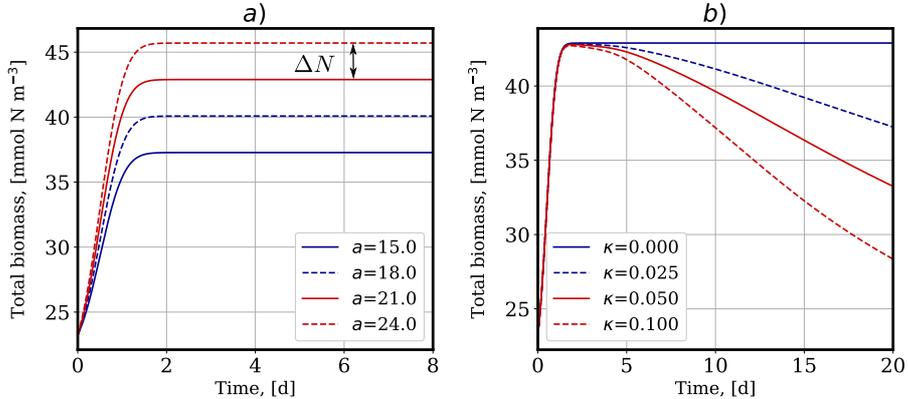}
			\caption{Time evolution of total biomass per unit volume in the euphotic zone.} 
			\label{NumericMassBalance}
		\end{center}
	\end{figure}
	
	In the first case we note that $a_{i+1}-a_i=3$ and therefore, the difference in total biomass per unit volume when the model is subjected to two pulses characterized by $a^{i+1}$ and $a^i$ is given by     
	\begin{equation}\label{NutrientIncrement}
	\Delta N=
	\int_{-\infty}^\infty\Big(\tn I_N^{i+1}(s)-\tn I_N^i(s)\Big)\,ds=
	(a_{i+1}-a_i)c\sqrt{2\pi}=3.64\quad \tn{mmol N m}^{-3},
	\end{equation}
	where the superscript $i$ in $\tn I_N^{i}(s)$ denotes an explicit dependence of the nutrient pulse on $a_i$. Additionally, since $\kappa=0$ no mass decay is expected to occur. This behavior is shown in Figure \ref{NumericMassBalance}-$(a)$.
	
	Figure \ref{NumericMassBalance}-$(b)$ shows that exact biomass conservation is obtained for $\kappa=0.0$ however, as long as $\kappa$ increases a higher rate of biomass decays is obtained due to detritus sinking. 
\end{subequations}
%
%
\subsection{Case study: a winter dinoflagellate bloom}\label{AlgalBloom}
\begin{figure}[!h]
	\begin{center}
		\includegraphics[width=\textwidth]{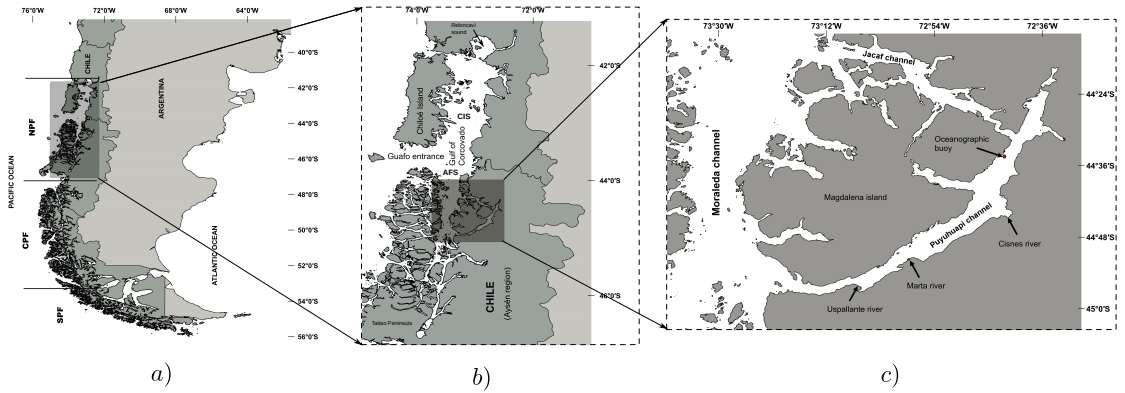}
		\caption{Geographical location. $a$) The Patagonian fjords area is located on the western coast of South America (41.5$^\circ$S-56$^\circ$S). $b$) Zoom view of the northern zone of the fjords area. $c$) A further zoom view allows to identify Puyuhuapi channel along with the location of the oceanographic buoy. The Puyuhuapi channel stretches from the Puyuhuapi village (44$^\circ$19'S,72$^\circ$33'W) to Moraleda channel (44$^\circ$57'S,73$^\circ$21'W) which connects it to open seas. Puyuhuapi Channel receives freshwater discharges from the Cisnes River (218 m$^3$s$^{-1}$ annual mean river flow).} 
		\label{PuyuhuapiLocation}
	\end{center}
\end{figure}

High rates of gross primary production and chlorophyll-$a$ concentration associated to an unusual winter bloom of dinoflagellates were reported at a fixed station in Puyuhuapi channel (Chilean Patagonia, 44$^\circ$35.30'S; 72$^\circ$43.60'W) during July 2015 \cite{MonteroEtAl2}. See Figure \ref{PuyuhuapiLocation} for location details. In the same article a detailed description of a winter sampling campaign from July 10 to 16, 2015 is provided. Collected data included hydrographic profiles and water samples from different depths. Field works were complemented with continuously recorded oceanographic and meteorological data at a nearby buoy.

A possible explanation for the bloom formation was proposed in \cite[$\S$4]{MonteroEtAl2}: high intensity winds associated to the passage of low pressure systems, intensified the water column's mixing processes, which induced the nutrient entrainment from deeper water levels into the euphotic zone. Once turbulent mixing decreased, re-stratification had place due to strong fresh water discharges from Cisnes river. As result, a nutrient enriched and stable euphotic layer was obtained. In this scenario and in spite of the low surface irradiance levels characteristic of austral winter, an algal bloom occurred. A hypothetical explanation for this phenomenon can be stated by linking the superior swimming skills of dinoflagellates, (which allows them to move to more advantageous zones in the water column) with a general warmer temperature of sea water. Both factors may have contributed decisively to provide appropriate comfort conditions for the bloom. 

In following we provide a detailed step-by-step procedure to apply the proposed two-layer model to a dinoflagellates bloom. Both, geometric dimensions and external forcing factors are built based on field's observations. Model's parameters were calibrated using GA-based methodology as explained in $\S$\ref{ParameterOptimizationGA}. It should be noted that most of the experimental data required by the model were taken from \cite{MonteroEtAl2} and therefore, they will not be repeated here.
%
%
\subsubsection{Two-layer model}\label{TwoLayerExample}
\begin{center}
	\begin{figure}[!h]
		\includegraphics[width=1\textwidth]{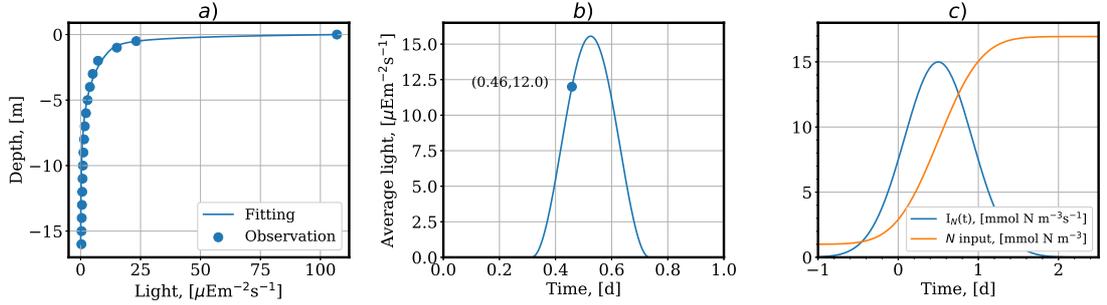}
		\caption{$a$) Solid circles corresponds to time-averaged values ​​of five experimentally obtained vertical profiles of underwater irradiance (days: July, 10, 12, 14 and 16, 2015 at 11 a.m.). The solid line corresponds to a fitting model for the mean values as given in \eqref{FittingL}. $b$) This figure shows the approximation used for the time evolution of the mean PAR in the upper layer. The curve is appropriately scaled in order to get $\tn I(0.46)=12.00$ [$\mu$Em$^{-2}$s$^{-1}$]. $c$) The solid-blue line corresponds to the time-dependent Gaussian pulse $\tn I_N(t)$ ($a=15.0$, $b=0.5$ and $c=0.424$) and the solid-orange line to the  curve defined in equation \eqref{IntPulse}.} 
		\label{Light}
	\end{figure}
\end{center}

\begin{subequations}
	The first step consist in determining the upper layer's depth which corresponds to the depth of the euphotic zone. We proceed as follows:
	\begin{enumerate}
		\item[i)] A mean vertical profile of light attenuation with depth was constructed by computing the time-averaged values ​​of five experimentally obtained vertical profiles of underwater irradiance measured in terms of photo-synthetically active radiation (PAR). Days: July, 10, 12, 14 and 16, 2015 at 11 a.m; see Figure \ref{Light}-$(a)$.
		\item[ii)] A fitting model for the values of mean vertical profile is given by
		\begin{equation}\label{FittingL}
		\Upsilon(d) = 87.56\,\exp(-4.881|d|)+19.31\,\exp(-0.3952|d|),
		\end{equation}
		where $d\in[-16.0,0.0]$ is a downward vertical coordinate for depth. 
		\item[iii)]	We consider that the euphotic zone extends from sea surface $d=0$ until a depth $d^\ast$ at which the following condition holds,
		\begin{equation}
		\Upsilon(d^\ast)/\Upsilon(0)=0.025.
		\end{equation}
		A simple calculation shows that $d^\ast\approx -5$ [m]. Therefore the first layer's depth is defined as the depth to which the 2.5\% of surface's PAR remains active.  
	\end{enumerate}
\end{subequations}
%
%
%
\subsubsection{Light availability}\label{solar_irradiance}
\begin{subequations}
	In this section we compute an analytical expression for light availability in the first layer. On the one hand, we note that mean value of photo-synthetically available radiation in the first layer at $t=0.46$ [d] (11 a.m.) is given by
	\begin{equation}\label{AverageLightL1}
	A=\dfrac{1}{|d^\ast|}\int_{d^\ast}^{0}\Upsilon(s)ds
	\approx 12.0\,\,[\mu\tn{Em}^{-2}\tn{s}^{-1}].
	\end{equation} 	
	Moreover, based on meteorological data we state that solar irradiance at the sea surface is described by
	\begin{equation}\label{Par1}
	\tn I(t)=
	\left\{
	\begin{array}{ll}
	\dfrac S 2\left(\sin\left(\dfrac{100\pi t}{21}-2\pi\right)+1\right), & \quad\tn{if }0.31\geqslant t\geqslant 0.73\,[d]\\
	0, & \quad\tn{otherwise}
	\end{array}
	\right.,
	\end{equation}
	where $S\approx100$ $[\mu\tn{Em}^{-2}\tn{s}^{-1}]$.
	
	In this work we assume that the time evolution of the mean PAR in the upper layer follows \eqref{Par1} but replacing $S=100$ by $S=15.5586=12.0/\tn I(0.4600)$. See Figure \ref{Light}-$(b)$.
\end{subequations}
%
%
\subsubsection{Nutrient input due to vertical mixing}\label{Nutrient_input_1}
\begin{subequations}
	As described in $\S$\ref{S.1} we focus on brief algal blooms triggered by intense climatic events involving high intensity winds and rainfalls. Accordingly, we consider a one day long idealized vertical mixing process, represented by $\tn I_N(t)$, that reaches its maximum intensity by the half of the day. The characteristics of the Gaussian pulse used to represent the vertical mixing are controlled by the scalars $a$, $b$ and $c$ as it is shown in \eqref{GaussianF1}. 
	
	We take $b=0.5$ since $\tn I_N(t)$ is assumed to be centered in the first day. To determine $c$ we assume a \textit{full width at half maximum}, $F_c$ , equal to 1.0 [d] along with the identity $F_c=2\sqrt{2\ln 2}c=1$ from which we get $c=0.424$. Additionally, $a$ is computed from
	\begin{equation}
	ac\sqrt{2\pi}=\int_{-\infty}^\infty\tn I_N(s)\,ds=\widehat{\tn N},
	\end{equation}
	where $\widehat{\tn N}$ is the total nutrient increment in the first layer due to vertical mixing. Taking $\widehat{\tn N}\approx 16.00$ yields to $a\approx 15.00$. Figure \ref{Light}-$(c)$ show the corresponding Gaussian pulse along with the curve
	\begin{equation}\label{IntPulse}
	\widehat{\tn N}(t)=\int_{-\infty}^t\tn I_N(s)\,ds.
	\end{equation}
\end{subequations}
which shows the time evolution of the nutrient input into the first layer. 
%
%
\subsubsection{Biogeochemical data}\label{BiogeochemicalData}
\begin{subequations}
	As it has been anticipated in $\S$\ref{MathModel}, for sake of model's consistency all the state variables are expressed in Nitrogen content per unit volume [mmol N m$^{-3}$]. Appropriate conversion units allow to express results in any other desired format. The procedure used to convert field data obtained in the sampling campaign from 10 to 16 July, 2015 is as follows,
	\begin{enumerate}[i)]
		\item \textit{Nutrient}: The NO$_3$ content was considered as the unique nutrients' source. It was taken from discrete observations at three depths: 2, 5 and 15 [m].
		\item \textit{Phytoplankton}: Field measurements of Chlorophyll-$a$ (Chl-$a$) concentration were used as a proxy for phytoplankton biomass. Two information sources were considered: i) surface's Chl-$a$ recorded using a multi-parameter water quality data collection system and, ii) vertical profiles constructed from Chl-$a$ samples taken at three depths (2, 5 and 15 [m]). Following \cite{BodeEtAl1}, Chlorophyll-$a$ to Carbon conversion was carried out with a C:Chl-$a$ ratio of 50 and a C:N ratio of 7.6 was used to convert Carbon to Nitrogen content.
		\item \textit{Zooplankton}: Plankton abundance was taken from observations at three depths (2, 5 and 15 [m]). The Carbon content associated to zooplankton abundance was estimated assuming an average value of 56 [$\mu$g C] for each individual \cite{EscribanoRodriguez1}. Then, a C:N ratio of 8.6 was applied to compute the nitrogen content \cite{Gismervik1}.
		\item \textit{Detritus}: As explained in $\S$\ref{ConceptualM} this state variable takes into account for POM plus DOM concentrations. However, in marine environments DOM plays a dominant role in ecosystem dynamics \cite{Thornton1}. In this work dissolved organic Carbon (DOC) was used as a proxy for autochthonous DOM \cite{HarveyKratzerAndersson1}. No data were available for POM. Additionally, a DOC:DON ratio equal to 10.4 was employed \cite{MarkagerEtAl1}.  
	\end{enumerate}
	Since no data were collected on July 8, surface concentration of Chl-$a$ was obtained by extrapolation from data collected by July 10 to 16. Nitrogen content was estimated from winter's measurements of previous years. No data were available for neither zooplankton nor DOC by this date. Finally, mean values for the state variables in the euphotic zone were estimated according to \eqref{AverageLightL1} by replacing $\Upsilon(d)$ by an appropriate interpolation function. See Table \ref{FieldDataConverted}. Moreover, Figure \ref{Data_interpolation} shows the mean values of $N$, $P$ and $Z$ obtained from field data along with their interpolation polynomials. The curve for detritus is not depicted in this figure since we assume an constant value $D=20.631$ [mmol N m$^{-3}$].    
	\begin{table}[!h]
		\centering
		\begin{tabular}{ |l|l|l|l|l|l|l|} 
			\hline
			Date              & July 6  & July 8 &  July 10 &  July 12  &  July 14 & July 16\\
			\hline
			Time, [d]         & 0.0     & 1.5    &  3.5     &  5.5      & 7.5      & 9.5\\
			Nutrient          & 1.000   & 11.200 &  5.827   &  2.181    & 1.831    & 3.752\\
			Phytoplankton     & 1.500   & 2.642  &  5.908   &  3.439    & 3.135    & 5.191\\
			Zooplankton       & 0.100   & --     &  0.788   & 4.871     & 1.484    & 0.469\\
			Detritus          & 20.631  & --     & 15.899   &  28.314   & 12.882   & 25.429\\
			\hline
		\end{tabular}
		\caption{Mean values of the state variables in the euphotic zone ([mmol N m$^{-3}$]). The first column corresponds to model's initial conditions.}
		\label{FieldDataConverted} 
	\end{table}

	The first column of Table \ref{FieldDataConverted} corresponds to the model's initial conditions. Those values were assigned taking into account a typical pre-bloom situation.  
	
	\begin{figure}[!h]
		\begin{center}
			\includegraphics[width=1\textwidth]{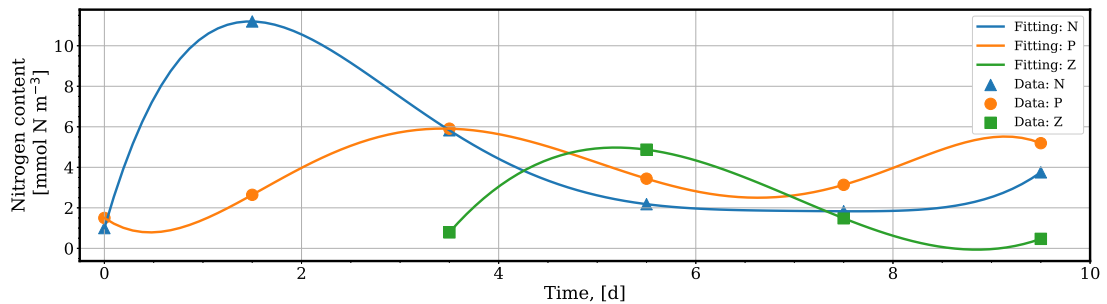}
			\caption{Mean values of the state variables $N$, $P$ and $Z$ in the euphotic zone obtained from field data. A constant value $D=20.631$ is assumed for detritus. Time continuous approximations of the state variables are constructed by means of polynomial fitting. Thus, we have: $N(t)=1.0+16.9453\,t-9.28\,t^2+1.93\,t^3-0.18\,t^4+0.0063\,t^5$, $P(t)=1.5-3.34\,t+4.50\,t^2-1.42\,t^3+0.17\,t^4-0.007\,t^5$ and $Z(t)=-53.93+28.23\,t-4.32\,t^2+0.205\,t^3$.}
			\label{Data_interpolation}
		\end{center}	
	\end{figure}
\end{subequations}
%
%
\subsubsection{Parameters calibration}
\begin{subequations}
	Model's parameters are calibrated considering the fitness function \eqref{Opt1} with weighting factors 
	\begin{equation}
	\left\{w_i\right\}_{i=1}^4=\{0.10, 0.40, 0.49, 0.01\},
	\end{equation}
	since we are interested in simulating high rates of primary production due to a wind driven bloom of dinoflagellates. The time interval of interest was $[0.0,9.0]$ [d] and 100 time steps were used during the simulations. 
	
	An optimization program was constructed in Fortran-2003 language that included an object-oriented\footnote{Download site: https://github.com/jacobwilliams/pikaia.} version of the open-source GA-based optimization subroutine Pikaia \cite{CharbonneauKnappReport1}. Simulations were carried out considering: i) an initial population composed by 1000 individuals \ie chromosomes encoding different values of $\bm\theta$ \eqref{Parameters1}; the initial population corresponds to a random sampling of $\ca S$, ii) the evolutionary process evolves until a converge tolerance of 10$^{-6}$ or a maximum of 10.000 generations is attained, iii) a cross-over probability equal to $p_{cross}=0.95$, iv) an initial vale for mutation rate $p_{mut}=0.005$ (adjustable depending on fitness) and v) a reproduction plan where the worst individual is always replaced. See \cite{CharbonneauKnappReport1,SivanandamDeepaBook1} for more details about GA-based algorithms.

	\begin{figure}[!h]
		\begin{flushleft}
			\includegraphics[width=1\textwidth]{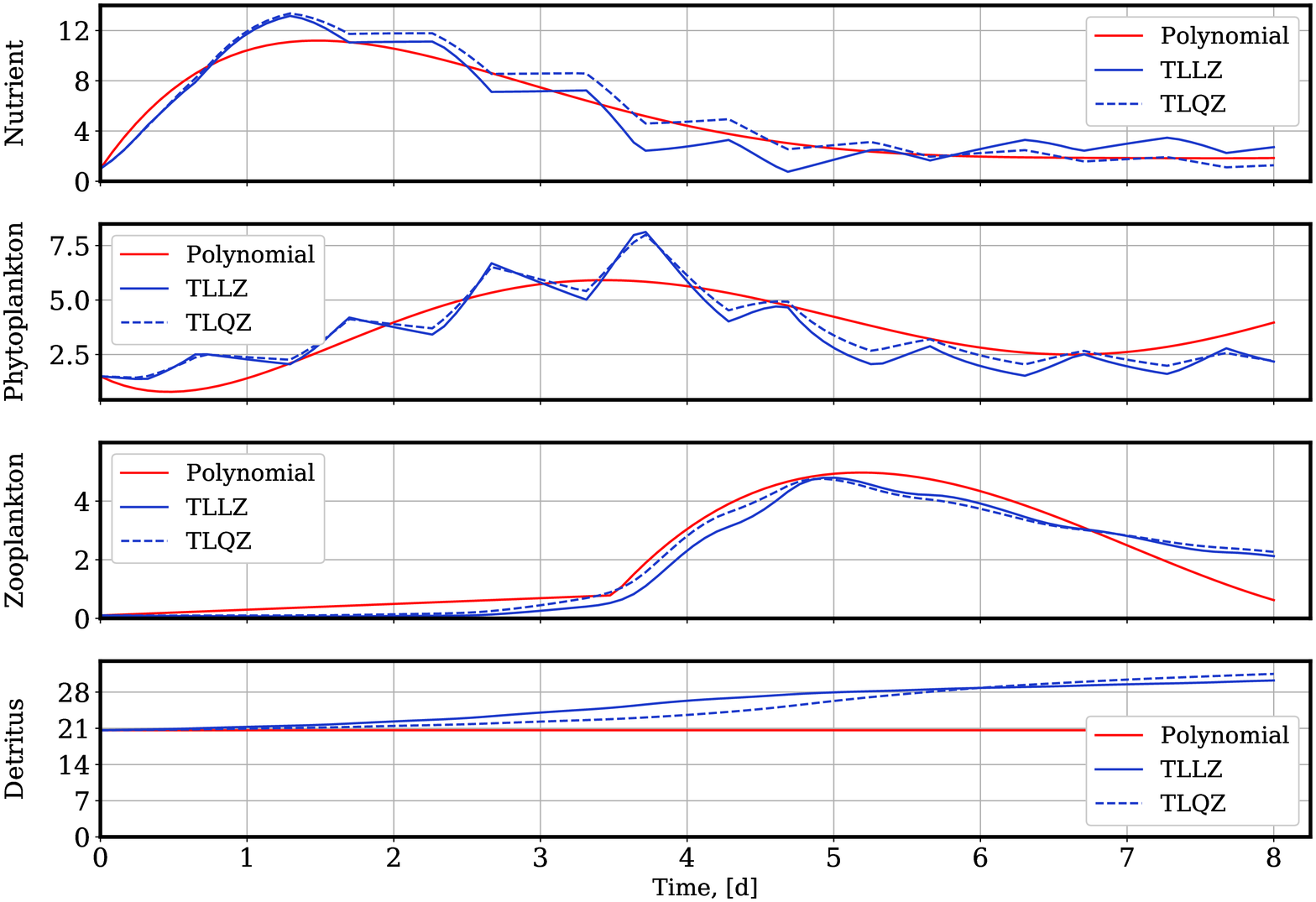}
			\caption{This figure presents a comparison between polynomial interpolations of the field data (solid red lines) and the numerical simulations corresponding to the cases: 1) TLLZ (solid blue line) and TLQZ (dashed blue line). A better fitting is obtained for TLQZ. Variables are expressed in [mmol N m$^{-3}$].}
			\label{Data_interpolation1}
		\end{flushleft}	
	\end{figure}

	Since there are no rules to follow in order to perform an ideal model calibration, several numerical tests were carried out revealing that some of the model's parameters are highly sensitive to variations in both, GA-parameters (\eg size of initial population, number of iterations, convergence's tolerance, $p_{cross}$, $p_{mut}$, etc.) and model forcing factors (\eg Gaussian pulse's parameters, analytical form of light availability, etc). Thus, the following four cases were considered:
	\begin{enumerate}[i)]
		\item TLLZ: This case considers that the daily available light evolves according to \eqref{Par1} and a linear zooplankton loss rate (\ie $\phi_z^\ast=0$). 
		\item TLQZ: Daily available light follows \eqref{Par1} and a quadratic zooplankton loss rate is considered (\ie $\phi_z^\ast\ne 0$). 
		\item MLLZ: This case considers that available light corresponds to the mean value $\int_0^1\tn I(s)\,ds=3.27$ [$\mu E$m$^{-2}$s$^{-1}$] along with a linear zooplankton loss rate.
		\item MLQZ: Mean available light equal to $3.27$ [$\mu E$m$^{-2}$s$^{-1}$] and quadratic zooplankton loss rate.       
	\end{enumerate}
	
	\begin{table}	
		\begin{centering}
			\begin{tabular}{|l|ccccccc|}
				\hline
				Situation & $k_N$ & $k_I$ & $\mu_m$ & $\phi_z$ & $\gamma_m$ & $\phi_p$ & $\epsilon$\\
				\hline
				TLLZ  & 0.00459 & 0.04515 & 1.07718 & 0.33840 & 0.00003 & 0.17056 & 0.03995\\
				TLQZ  & 0.86336 & 0.05112 & 0.94848 & 0.10830 & 0.00005 & 0.08091 & 0.02791\\
				MLLZ  & 0.00138 & 9.19370 & 1.93210 & 0.46930 & 0.01245 & 0.25602 & 0.04509\\
				MLQZ  & 0.12540 & 11.1625 & 2.62672 & 0.24589 & 0.03842 & 0.31866 & 0.03423\\                        
				\hline
			\end{tabular}
			\caption{Model's parameters (part 1). }
			\label{OptimalParameters1}
		\end{centering}
	\end{table}

	\begin{table}	
		\begin{centering}
			\begin{tabular}{|l|cccc|c|}
				\hline
				Situation & $g$ & $\beta$ & $\phi_z^\ast$ & $\kappa$ & $\Gamma(\bm\theta)$\\
				\hline
				TLLZ & 17.8036 & 0.97842 & --      & -- & -62.92405\\
				TLQZ & 26.8129 & 0.99702 & 0.05820 & -- & -45.69100\\
				MLLZ & 17.7468 & 0.99007 & --      & -- & -105.6393\\
				MLQZ & 29.7383 & 0.99671 & 0.07773 & -- & -95.78919\\                        
				\hline
			\end{tabular}
			\caption{Model's parameters (continuation).}
			\label{OptimalParameters2}
		\end{centering}
	\end{table}
	
	Tables \ref{OptimalParameters1} and \ref{OptimalParameters2} present the set of optimal parameters for every case along with the corresponding values of the fitness function $\Gamma(\bm\theta)$. Figures \ref{Data_interpolation1} and \ref{Data_interpolation2} show that a better agreement between experimental data and model simultation is obtained for TLQZ ($\Gamma=-45.69100$) when compared with TLLZ ($\Gamma=-62.92405$), and the same applies for MLQZ ($\Gamma=-95.78919$) when compared with MLLZ ($\Gamma=-105.6393$). These results provide numerical evidence indicating that including the quadratic term $\phi_z^\ast (Z(t))^2$ contributes to improve the model's ability to represent the system's dynamics. In a wider context the best fitting is obtained for TLQZ indicating that a detailed representation of the external forcing factors is beneficial for the model's reliability.

	\begin{figure}[!h]
		\begin{center}
			\includegraphics[width=\textwidth]{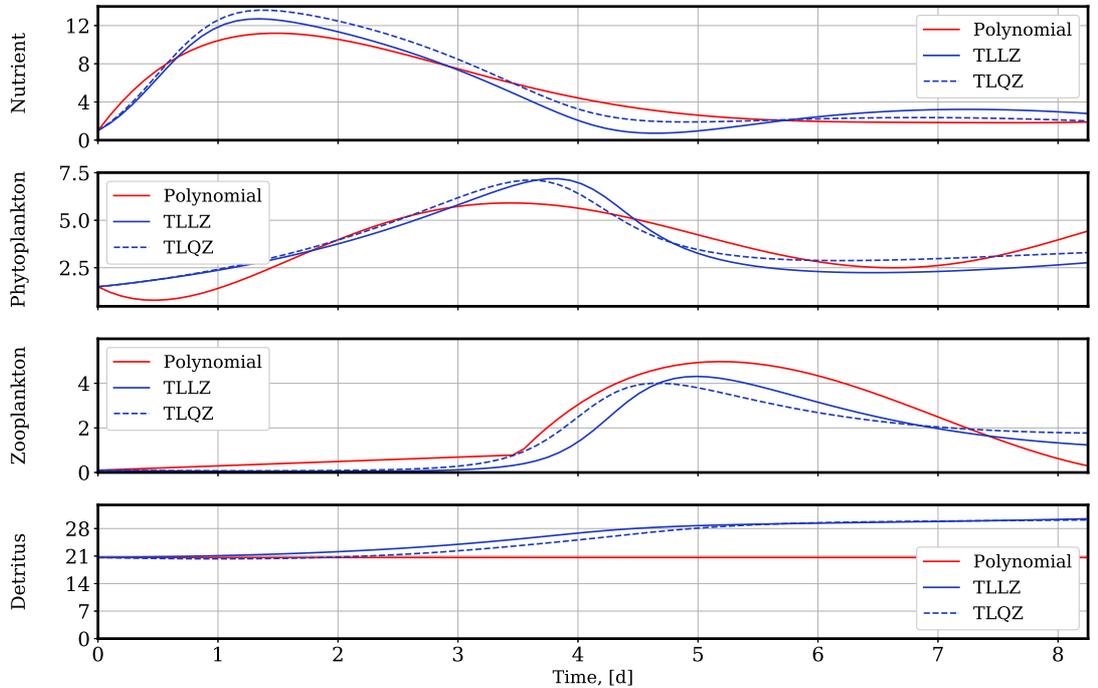}
			\caption{This figure presents a comparison between polynomial interpolations of the field data (solid red lines) and the numerical simulations corresponding to the cases: 1) MLLZ (solid blue line) and MLQZ (dashed blue line). A better fitting is obtained for MLQZ. Variables are expressed in [mmol N m$^{-3}$].}
			\label{Data_interpolation2}
		\end{center}	
	\end{figure}

	An inspection of Tables \ref{OptimalParameters1} and \ref{OptimalParameters2} reveals that:
	\begin{enumerate}[i)]
		\item The parameter space $\ca S$ shows a complex topology with multiple local maxima. The GA used for calibration evidences this feature since the inclusion of a quadratic term in TLLZ to get TLQZ yields to significant variations in some parameters (\eg $k_N=0.00459\rightarrow0.86336$; $\phi_p=0.17056\rightarrow0.08091$; $\phi_z=0.33840\rightarrow0.10830$). This applies to MLLZ and MLQZ as well. We hypothesize that this behavior is related to size of the initial population (1000 chromosomes), since a simple calculation shows that if we admit 10 possible values for every parameter, we get $10^{10}$ combinations that are possible solutions. Thus a relatively small initial sampling of $\ca S$ could limit the evolution paths of the GA, confining the searching process to certain local maxima. Several tests carried out with bigger initial populations yielded to analogous results.
		\item The form adopted for the external forcing factors strongly influences the set of optimal parameters. This is evidenced by comparing the set of parameters for TLLZ and MLLZ (\eg $k_I=0.04515\rightarrow9.19370$; $\gamma_m=0.00003\rightarrow0.01245$). The same applies to TLQZ and MLQZ as well. In spite of these differences, a reasonable agreement between the models and experimental data is obtained in all the cases. This situation indicates that we are probably facing with a problem that admits multiple approximated solutions. See Figures \ref{Data_interpolation1} and \ref{Data_interpolation2}.
		\item In spite of the above observations, some parameters' values remain remarkably stable across the simulations; see \eg $\mu_m$, $\epsilon$, $g$, $\beta$ ad $\phi_z^\ast$.      
	\end{enumerate}
	As closure comment we state that in spite it was not possible to find a unique set of optimal parameters, it is possible to find certain combinations that fit reasonably well to experimental data.   
\end{subequations}
%
%
\subsubsection{Ecosystem response to a sequence of nutrient pulses}
\begin{subequations}
	This example focuses on simulating a sequence of intermittent peaks of primary production associated to (winter) dinoflagellates' blooms driven by nutrient pulses. To this end we consider the model described in $\S$\ref{TwoLayerExample} to \ref{BiogeochemicalData}. Model's parameter are those of the TLQZ case in Tables \ref{OptimalParameters1} and \ref{OptimalParameters2} excepting for $\gamma_m=1\times10^{-4}$ and $\beta=0.75$. The sequence of nutrient pulses is shown in Figure \ref{TimeHistoy1}-$(c)$.

	Figure \ref{TimeHistoy1}-$(a)$ shows the time history response of the state variables $N$, $P$ and $Z$. Every pulse triggers a peak of nutrients in the euphotic zone followed by a phytoplankton peak approximately three days later. A third peak corresponding to zooplankton has place approximately two days after the second one. The sequence of peaks $N\rightarrow P\rightarrow Z$ accounts for the biomass flow through the trophic web. Nutrient, phytoplankton and zooplankton depletion follows after every pulse episode due to conversion to detritus which acts as a long term reservoir of organic Carbon; see Figure \ref{TimeHistoy1}-$(b)$. Therefore, the proposed model is able to reproduce the intermittent sequence of $N$-$P$-$Z$ peaks which are followed by depletion's scenarios as reported for dinoflagellates' blooms \cite{MonteroEtAl2}.    
	\begin{figure}[!h]
		\begin{center}
			\includegraphics[width=\textwidth]{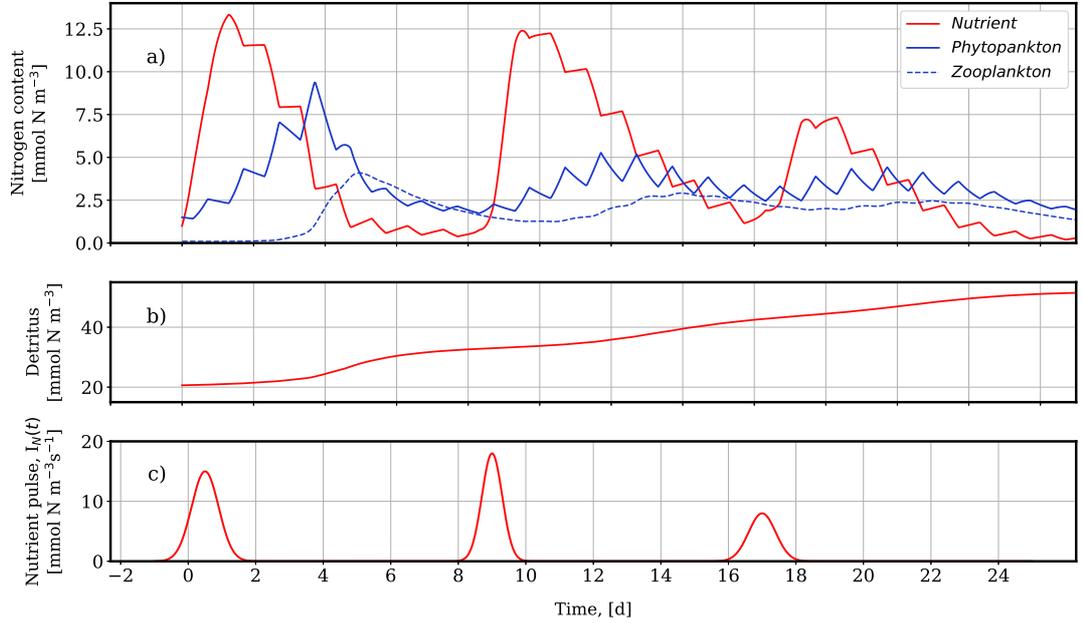}
			\caption{Two-layer model response to a sequence of wind driven mixing processes. $a)$ Time evolution of nutrient, phytoplankton and zooplankton content per unit volume. In every state variable it is possible to appreciate three peaks followed by depletion due to biomass transfer to detritus. $b)$ Time evolution of detritus content per unit volume. $c)$ Time history of external forcing. Three nutrient pulses are characterized by: 1) $a=15.0$, $b=0.5$, $c=0.4$; 2) $a=18.0$, $b=9.0$, $c=0.3$; and 3) $a=8.0$, $b=17.0$, $c=0.4$; see \eqref{GaussianF1}.}\label{TimeHistoy1}
		\end{center}	
	\end{figure}
	
	Primary production at time $t\in[0.0,25.0]$, $PP(t)$, is computed as
	\begin{equation}\label{PP1}
	PP(t)=\int_0^tJ(N(s),\tn I(s))P(s)\,ds,
	\end{equation}
	which is approximated by the trapezoidal rule according to
	\begin{equation}\label{PP1numerics}
	PP^k=\frac h 2\sum_{i=2}^k\Big[J\left(N^{i-1},\tn I(t^{i-1})\right)P^{i-1}+
	J\left(N^{i},\tn I(t^{i})\right)P^{i}\Big]\approx PP(t^k).
	\end{equation}
	The same procedure applies to compute the \textit{total phytoplankton grazing} but replacing $J(N,\tn I)P$ in \eqref{PP1} by $G(\epsilon,g,P)Z$. In fact the time integral of any other flux's rate of those shown in Figure 2 provides a measure of the corresponding transfer of biomass between functional groups. 
	
	Figure \ref{TimeHistoy2}-$(a)$ shows the time evolution of primary production rate, $J(N,I)P$, (PP-rate) and phytoplankton grazing rate, $G(\epsilon,g,P)Z$, (GP-rate). Due to the discontinuous nature of the light availability function \eqref{Par1}, the PP-rate curve corresponds to an intermittent sequence of active (light-available) periods and non-active (darkness) periods. The corresponding envelope curve presents three characteristic peaks associated to the maxima in phytoplankton content. The GP-rate curve is continuous in time and it is also possible to identify three peaks associated to the maxima in zooplankton production.

	\begin{figure}[!h]
		\begin{center}
			\includegraphics[width=\textwidth]{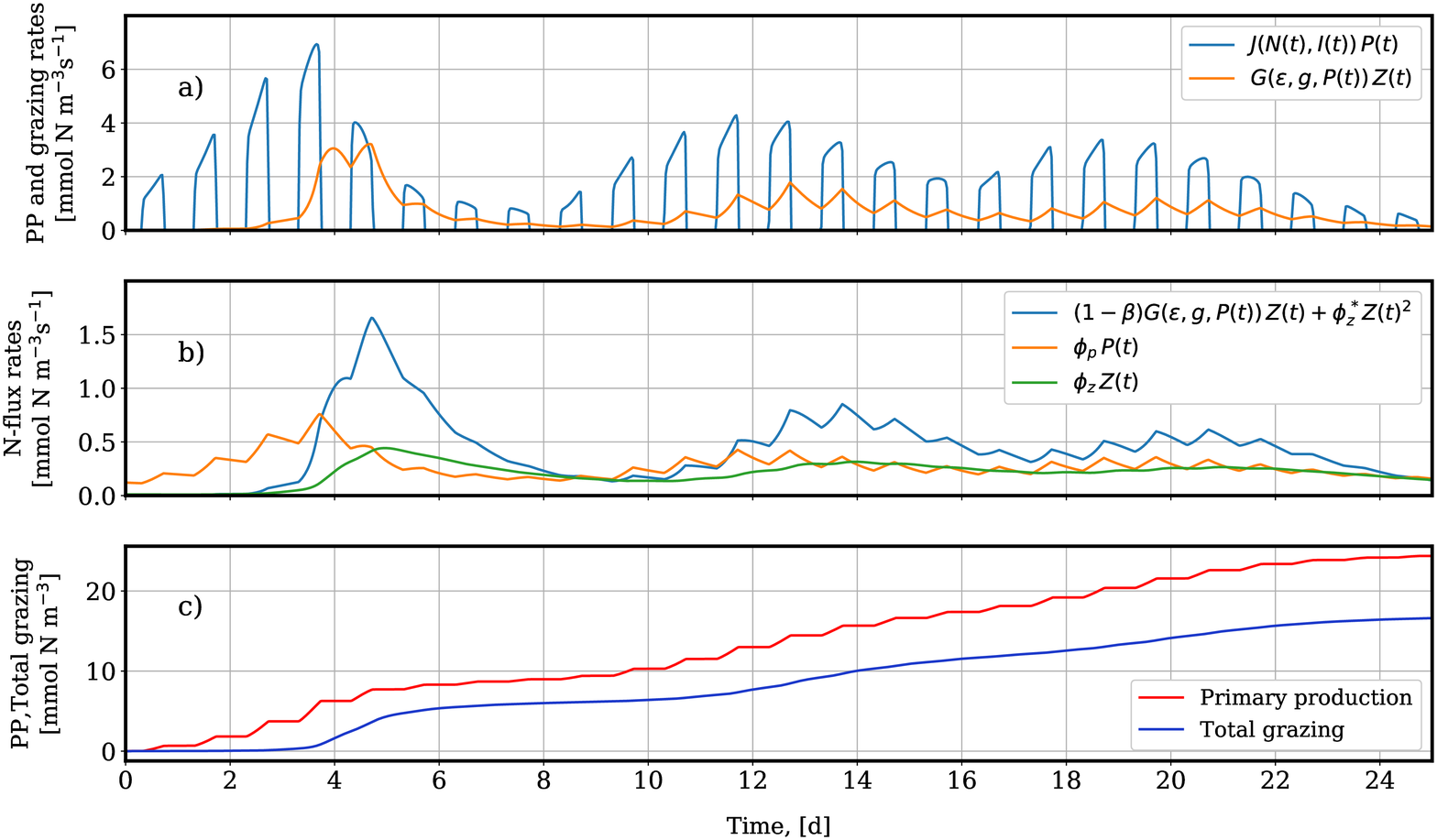}
			\caption{$a)$ Time history response of PP- and GP-rates. In the first case, the flux rate is limited by light availability thus producing an intermittent sequence of active and inactive periods. The corresponding envelope-curve still retains the three-peaks shape associated to the nutrient to phytoplankton conversion. On he contrary, time evolution of the grazing rate is continuous but also presents a three-peaks shape. $b)$ Time evolution of zooplankton to detritus conversion rate, phytoplankton mortality rate and zooplankton excretion rate. $c)$ Time evolution of primary production and total phytoplankton grazing. The value of $PP(25)$ is $24.38$ [mmol N m$^{-3}$].}\label{TimeHistoy2}
		\end{center}	
	\end{figure}
	In Figure \ref{TimeHistoy2}-$(b)$ the time history of zooplankton to detritus conversion rate, $(1-\beta)G(\epsilon,g,P)Z+\phi_z^\ast Z^2$ (ZD-rate), phytoplankton mortality rate, $\phi_pP$, and zooplankton excretion rate, $\phi_z Z$, are shown. In every one of these cases, a three-peaks shape curve can be identified and it is possible to see that the ZD-rate dominates the detritus production process.        Figure \ref{TimeHistoy2}-$(c)$ presents the primary production and the total phytoplankton grazing curves (per unit volume) obtained with \eqref{PP1numerics}. In what regards to applications, the PP curve allows to make estimations of biomass production in a marine water body. For example, under the assumption that the above results represent the average ecosystem behavior in an area of 1 km$^{2}$, allow to compute a value of phytoplankton production by the day 25 equals to $24.38\times 5\times10^6=121.9\times10^6$ [mmol N]. A straightforward conversion can be used to express it in carbon or biomass units.     
\end{subequations}
%
%
%
%
\section{Conclusions and further research}\label{Conclusions}
A novel mathematical model for brief phytoplankton blooms occurring in the euphotic zone of semi-enclosed marine bodies located in the fjord's area of western Patagonia has been proposed and validated in a realistic case. The model presents an appropriate balance between complexity and applicability that allows its use for case studies in marine zones where field-data sparsity discourages the application of sophisticated computer models. A two-layer description of a representative volume of the water column is considered. The first layer corresponds to the euphotic layer where a biogeochemical model of NPZD-type is used to simulate a mass-conserving trophic web. Wind driven turbulent mixing of the water column introduces upward pulses of nutrient in the euphotic zone that trigger algal bloom events by means of the photosynthetic production of phytoplankton. Biomass flows through the food web appear as a consequence of zooplankton grazing. Moreover, the mortality and excretion coming from living organisms contribute to increase the detritus's content. Time-dependent Gaussian pulses are used to describe the nutrients' entrainment. Biomass losses are taken into account by means of detritus sinking to lower levels in the water column. In this manner, the total biomass is no longer invariant. Therefore, the ecosystem dynamics is described by strictly positive trajectories that correspond to the solution of an externally forced, non-autonomous system of ODE's. A three-stages time integration method based on the so called splitting-composition techniques is formulated to solve the system. The proposed algorithm produce discrete trajectories that are strictly positives along with providing an exact estimations of the biomass balance during algal blooms. Model calibration is carried out by applying a genetic algorithm-based optimization procedure. Model's parameters are adjusted following an evolutionary process in order to assimilate specific field data. 

The proposed model was applied to the study of an infrequent winter bloom of dinoflagellates in an austral fjord (2015) \cite{MonteroEtAl2}. Both the geometric and biogeochemial properties of the two-layer model were determined from observational data collected in field works as well as data recorded in a nearby oceanographic buoy. Procedures to define the time dependent Gaussian pulse of nutrients along with the daily light availability were also provided. Model's calibration deserves some remarks since it resulted to be a complex task due to the relatively high number of model's parameters ($\approx 10$). The GA-based optimization searches revealed a topologically intricate parameters space involving multiple local maxima, which prevented obtaining a single set of optimal parameters for the problem. Instead four sets of (quasi-) optimal parameters were provided for representative case studies. A simplified sensitivity analysis was carried out identifying the most stable parameters. Finally, the a calibrated version of the proposed model was successfully applied for simulating the time history response of the ecosystem when is subjected to a sequence of three wind induced nutrient pulses. The numerical simulations allow to reproduce an scenario where an intermittent sequence of primary production peaks was followed by a coherent flow of biomass through the food-web's functional groups. It was also possible to provide some rough estimations of biomass production (per km$^2$) due to algal blooms.

Future research should be oriented to: i) improve the quality of field data in order to implement more reliable data assimilation procedures, ii) to extend the proposed model to study the seasonal cycle of primary production observed in marine bodies of western Patagonia and iii) to couple the proposed model to an one-dimensional hydrodynamic water column model in order to generate more sophisticated representations of the interaction between biogeochemistry and turbulent mixing processes in the ocean.               
%
%
\section*{Acknowledgments}

This research was funded by the Fondo de Investigaci\'on para la Competitividad, 2018, BIP 40000236-0 and the Programa CONICYT-Chile R17A10002. We also thank Paulina Montero for providing some data used in $\S$\ref{AlgalBloom}. This support is gratefully acknowledged.
%
%
\bibliographystyle{unsrt}
\bibliography{Biblio}

\begin{thebibliography}{10}

\bibitem{BurchardEtAl2}
H.~Burchard, E.~Deleersnijder, and A.~Meister.
\newblock A high-order conservative {P}atankar-type discretisation for stiff
  systems of production-destruction equations.
\newblock {\em Applied Numerical Mathematics}, 47:1--30, 2003.

\bibitem{MonteroEtAl2}
P.~Montero, I.~P\'erez-Santos, G.~Daneri, M.H. Guti\'errez, G.~Igor, R.~Seguel,
  D.~Purdie, and D.W. Crawford.
\newblock A winter dinoflagellate bloom drives high rates of primary production
  in a {P}atagonian fjord ecosystem.
\newblock {\em Estuarine, Coastal and Shelf Science}, 199:105--116, 2017.

\bibitem{IriarteEtAl1}
J.L. Iriarte, H.E. Gonz\'alez, and L.~Nahuelhual.
\newblock Patagonian {F}jord {E}cosystems in southern {C}hile as a highly
  vulnerable region: Problems and needs.
\newblock {\em Ambio. A Journal of the Human Environment}, 39(7):463--466,
  2010.

\bibitem{Pickard1}
G.L. Pickard.
\newblock Some physical oceanographic features of inlets of {C}hile.
\newblock {\em Journal of the Fisheries Research Board of Canada},
  28(8):1077--1106, 1971.

\bibitem{SilvaVargas1}
N.~Silva and C.A. Vargas.
\newblock Hypoxia in chilean patagonian fjords.
\newblock {\em Progress in Oceanography}, 129(Part A):62--74, 2014.

\bibitem{HoweEtAl1}
J.A. Howe, W.E.N. Austin, M.~Forwick, M.~Paetzel, R.~Harland, and A.G. Cage.
\newblock Fjord systems and archives: a review.
\newblock {\em The Geological Society of London}, 344:5--15, 2010.

\bibitem{NeshybaFonseca1}
S.~Neshyba and T.~Fonseca.
\newblock Evidence for counterflow to the {W}est {W}ind {D}rift off {S}outh
  {A}merica.
\newblock {\em Journal of Geophysical Research}, 85(C9):4888--4892, 1980.

\bibitem{Moffat1}
C.~Moffat.
\newblock Wind-driven modulation of warm water supply to a proglacial fjord,
  {J}orge {M}ontt {G}lacier, {P}atagonia.
\newblock {\em Geophysical Research Letters}, 41(11):3943--3950, 2014.

\bibitem{PantojaEtAl1}
S.~Pantoja, J.L. Iriarte, and G.~Daneri.
\newblock Oceanography of the {C}hilean {P}atagonia.
\newblock {\em Continental Shelf Research}, 31(3-4):149--153, 2011.

\bibitem{Romero1}
H.~Romero.
\newblock Geograf\'ia de los {C}limas. {G}eograf\'ia de {C}hile.
\newblock Instituto {G}eogr\'afico {M}ilitar. {T}omo {X}{I}, 243pp, 1985.

\bibitem{Stigebrandt1}
A.~Stigebrandt.
\newblock A mechanism governing the estuarine circulation in deep, strongly
  stratified fjords.
\newblock {\em Estuarine, Coastal and Shelf Science}, 13(2):197--211, 1981.

\bibitem{SchneiderEtAl1}
W.~Schneider, I.~P\'erez-Santos, L.~Ross, L.~Bravo, R.~Seguel, and
  F.~Hern\'andez.
\newblock On the hydrography of {P}uyuhuapi {C}hannel, {C}hilean {P}atagonia.
\newblock {\em Progress in Oceanography}, 129(Part A):8--18, 2014.

\bibitem{Sievers1}
H.~Sievers.
\newblock Temperature and salinity in the austral {C}hilean channels and
  fjords. {P}rogress in the oceanographic knowledge of {C}hilean interior
  waters, from {P}uerto {M}ontt to {C}ape {H}orn, pp. 31-36.
\newblock {C}omit\'e {O}ceanogr\'afico {N}acional -- {P}ontificia {U}niversidad
  {C}at\'olica de {V}alpara\'iso, {V}alpara\'iso, Chile, 2008.

\bibitem{DavilaEtAl1}
P.M. D\'avila, D.~Figueroa, and E.~M\"uller.
\newblock Freshwater input into the coastal ocean and its relation with the
  salinity distribution off austral {C}hile (35-55°{S}).
\newblock {\em Continental Shelf Research}, 22(3):521--534, 2002.

\bibitem{SilvaCalvete1}
N.~Silva and C.~Calvete.
\newblock Physical and chemical oceanographic features of southern chilean
  inlets between {P}enas {G}ulf and {M}agellan {S}trait ({C}imar-{F}iordo 2
  cruise).
\newblock {\em Ciencia y Tecnología del Mar}, 25(1):23--88, 2002.

\bibitem{PerezEtAl1}
I.~P\'erez-Santos, J.~Garc\'es-Vargas, W.~Schneider, L.~Ross, S.~Parra, and
  A.~Valle-Levinson.
\newblock Double-diffusive layering and mixing in {P}atagonian fjords.
\newblock {\em Progress in Oceanography}, 129(Part {A}):35--49, 2014.

\bibitem{CastilloValenzuela1}
M.~Castillo and C.~Valenzuela.
\newblock 4.2 {C}irculation regime in the austral {C}hilean channels and
  fjords.
\newblock Progress in the oceanographic knowledge of {C}hilean interior waters,
  from {P}uerto {M}ontt to {C}ape {H}orn. {N}. {S}ilva \& {S}. {P}alma (eds.).
  {C}omit\'e {O}ceanogr\'afico {N}acional - {P}ontificia {U}niversidad
  {C}at\'olica de {V}alpara\'iso, {V}alpara\'iso, Chile. Pp. 59-62., 2008.

\bibitem{RossEtAl1}
L.~Ross, A.~Valle-Levinson, I.~P\'erez-Santos, F.J. Tapia, and W.~Schneider.
\newblock Baroclinic annular variability of internal motions in a {P}atagonian
  fjord.
\newblock {\em Journal of Geophysical Research: Oceans}, 120(8):5668--5685,
  2015.

\bibitem{CastroEtAl1}
L.R. Castro, M.A. C\'aceres, N.~Silva, M.I.~Mun\ {o}z, R.~Le\'on, M.F.
  Landaeta, and S.~Soto-Mendoza.
\newblock Short-term variations in mesozooplankton, ichthyoplankton, and
  nutrients associated with semi-diurnal tides in a patagonian {G}ulf.
\newblock {\em Continental Shelf Research}, 31(3-4):282--292, 2011.

\bibitem{SilvaPalma1}
N.~Silva and S.~Palma.
\newblock 1.1 the {C}{I}{M}{A}{R} {P}rogram in the austral {C}hilean channels
  and fjords.
\newblock Progress in the oceanographic knowledge of {C}hilean interior waters,
  from {P}uerto {M}ontt to {C}ape {H}orn. {C}omit\'e {O}ceanogr\'afico
  {N}acional - {P}ontificia {U}niversidad {C}at\'olica de {V}alpara\'iso,
  {V}alpara\'iso, Chile. Pp. 11-15., 2008.

\bibitem{ArheimerNilssonLindstrom}
B.~Arheimer, J.~Nilsson, and G.~Lindstr\"om.
\newblock Experimenting with coupled hydro-ecological models to explore measure
  plans and water quality goals in a semi-enclosed {S}wedish bay.
\newblock {\em Water}, 7:3906--3924, 2015.

\bibitem{OuteiroVillasante1}
L.~Outeiro and S.~Villasante.
\newblock Linking salmon aquaculture synergies and trade-offs on ecosystem
  services to human wellbeing constituents.
\newblock {\em Ambio}, 42(8):1022--1036, 2013.

\bibitem{BartonFloysand1}
J.R. Barton and A.~Fløysand.
\newblock The political ecology of {C}hilean salmon aquaculture, 1982-2010: {A}
  trajectory from economic development to global sustainability.
\newblock {\em Global Environmental Change}, 20(4):739--752, 2010.

\bibitem{HosonoEtAl1}
A.~Hosono, M.~Iizuka, and J.~Katz, editors.
\newblock {\em Chile's Salmon Industry. {P}olicy Challenges in Managing Public
  Goods}.
\newblock Springer Japan, 2016.

\bibitem{BuschmannEtAl2}
A.H. Buschmann, F.~Cabello, K.~Young, J.~Carvajal, D.A. Varela, and
  L.~Henr\'iquez.
\newblock Salmon aquaculture and coastal ecosystem health in {C}hile:
  {A}nalysis of regulations, environmental impacts and bioremediation systems.
\newblock {\em Ocean \& Coastal Management}, 52(5):243--249, 2009.

\bibitem{StrubEtAl1}
P.T. Strub, J.M. Mesias, V.~Montecino, J.~Rutllant, and S.~Salinas.
\newblock Coastal ocean circulation off western south {A}merica.
\newblock In: {R}obinson, {C}., {B}rink, {T}. ({E}ds.), {T}he {S}ea. {J}ohn
  {W}iley \& {S}ons, {N}ew {Y}ork, pp. 273-313, 1998.

\bibitem{EscribanoEtAl1}
R.~Escribano, M.~Fern\'andez, and A.~Aran\'is.
\newblock Physical-chemical processes and patterns of diversity of the chilean
  eastern boundary pelagic and benthic marine ecosystems: {A}n overview.
\newblock {\em Gayana}, 67(2):190--205, 2003.
\newblock \url{http://dx.doi.org/10.4067/S0717-65382003000200008}.

\bibitem{IriarteEtAl2}
J.L. Iriarte, H.E. Gonz\'alez, K.K. Liu, C.~Rivas, and C.~Valenzuela.
\newblock Spatial and temporal variability of chlorophyll and primary
  productivity in surface waters of southern {C}hile (41.5--43$^\circ$s).
\newblock {\em Estuarine, Coastal and Shelf Science}, 74(3):471--480, 2007.

\bibitem{CastilloEtAl2}
M.I. Castillo, U.~Cifuentes, O.~Pizarro, L.~Djurfeldt, and M.~Caceres.
\newblock Seasonal hydrography and surface outflow in a fjord with a deep sill:
  the {R}eloncav\'i fjord, {C}hile.
\newblock {\em Ocean Science}, 12(2):533--544, 2016.

\bibitem{IriarteEtAl3}
J.L. Iriarte, S.~Pantoja, and G.~Daneri.
\newblock Oceanographic processes in {C}hilean fjords of {P}atagonia: {F}rom
  small to large-scale studies.
\newblock {\em Progress in Oceanography}, 129(Part {A}):1--7, 2014.

\bibitem{TorresEtAl2}
R.~Torres, N.~Silva, B.~Reid, and M.~Frangopulos.
\newblock Silicic acid enrichment of subantarctic surface water from
  continental inputs along the {P}atagonian archipelago interior sea
  (41-56$^\circ$s).
\newblock {\em Progress in Oceanography}, 129(Part {A}):50--61, 2014.

\bibitem{LafonEtAl1}
A.~Lafon, N.~Silva, and C.A. Vargas.
\newblock Contribution of allochthonous organic carbon across the {S}errano
  {R}iver {B}asin and the adjacent fjord system in {S}outhern {C}hilean
  {P}atagonia: {I}nsights from the combined use of stable isotope and fatty
  acid biomarkers.
\newblock {\em Progress in Oceanography}, 129(Part {A}):98--113, 2014.

\bibitem{BerdaletEtAl1}
E.~Berdalet, M.~Montresor, B.~Reguera, S.~Roy, H.~Yamazaki, A.~Cembella, and
  R.~Raine.
\newblock Harmful algal blooms in fjords, coastal embayments, and stratified
  systems: Recent progress and future research.
\newblock {\em Oceanography}, 30(1):46--57, 2017.

\bibitem{DiazEtAl1}
P.A. D\'iaz, G.~\'Alvarez, D.~Varela, I.~P\'erez-Santos, M.~D\'iaz, C.~Molinet,
  M.~Seguel, A.~Aguilera-Belmonte, L.~Guzm\'an, E.~Uribe, J.~Rengel,
  C.~Hern\'andez, C.~Segura, and R.I. Figueroa.
\newblock Impacts of harmful algal blooms on the aquaculture industry: {C}hile
  as a case study.
\newblock {\em Perspectives in Phycology}, 6(1-2):39--50, 2019.

\bibitem{TettPortillaGillibrandInall1}
P.~Tett, E.~Portilla, P.A. Gillibrand, and M.~Inall.
\newblock Carrying and assimilative capacities: the {A}{C}{E}x{R}-{L}{E}{S}{V}
  model for sea-loch aquaculture.
\newblock {\em Aquaculture Research}, 42:51--67, 2011.

\bibitem{JickellsEtAl1}
T.~Jickells, J.~Andrews, S.~Barnard, P.~Tett, and S.~van Leeuwen.
\newblock {\em Natural Sciences Modelling in Coastal and Shelf Seas}, volume~9.
\newblock Springer, 2015.

\bibitem{DoglioliEtAl1}
A.M. Doglioli, M.G. Magaldi, L.~Vezzulli, and S.~Tucci.
\newblock Development of a numerical model to study the dispersion of wastes
  coming from a marine fish farm in the {L}igurian {S}ea ({W}estern
  {M}editerranean).
\newblock {\em Aquaculture}, 231(1-4):215--235, 2004.

\bibitem{WanEtAl1}
D.~Wan, J.M. Klymak, M.G.G. Foreman, and S.F. Cross.
\newblock Barotropic tidal dynamics in a frictional subsidiary channel.
\newblock {\em Continental Shelf Research}, 105:101--111, 2015.

\bibitem{PonteCornuelle1}
A.L. Ponte and B.D. Cornuelle.
\newblock Coastal numerical modelling of tides: Sensitivity to domain size and
  remotely generated internal tide.
\newblock {\em Ocean Modelling}, 62:17--26, 2013.

\bibitem{HetlandDiMarco1}
R.D. Hetland and S.F. DiMarco.
\newblock Skill assessment of a hydrodynamic model of circulation over the
  {T}exas-{L}ouisiana continental shelf.
\newblock {\em Ocean Modelling}, 43-44:64--76, 2012.

\bibitem{GhilEtAl1}
M.~Ghil, Y.~Feliks, and L.U. Sushama.
\newblock Baroclinic and barotropic aspects of the wind-driven ocean
  circulation.
\newblock {\em Physica {D}: {N}onlinear {P}henomena}, 167(1-2):1--35, 2002.

\bibitem{SadrinasabKampf1}
M.~Sadrinasab and J.~K\"{a}mpf.
\newblock Three-dimensional flushing times of the {P}ersian {G}ulf.
\newblock {\em Geophysical Research Letters}, 31(24):1--4, 2004.

\bibitem{DelandmeterEtAl1}
P.~Delandmeter, S.E. Lewis, J.~Lambrechts, E.~Deleersnijder, V.~Legat, and
  E.~Wolanski.
\newblock The transport and fate of riverine fine sediment exported to a
  semi-open system.
\newblock {\em Estuarine, Coastal and Shelf Science}, 167(Part B):336--346,
  2015.

\bibitem{LuShi1}
L.~Lu and J.Z. Shi.
\newblock The dispersal processes within the tide-modulated {C}hangjiang
  {R}iver plume, {C}hina.
\newblock {\em International Journal for Numerical Methods in Fluids},
  55(12):1143--1155, 2007.

\bibitem{LacroixEtAl1}
G.~Lacroix, K.~Ruddick, J.~Ozer, and C.~Lancelot.
\newblock Modelling the impact of the {S}cheldt and {R}hine/{M}euse plumes on
  the salinity distribution in {B}elgian waters (southern {N}orth {S}ea).
\newblock {\em Journal of Sea Research}, 52(3):149--163, 2004.

\bibitem{PortillaTettGillibrandInall1}
E.~Portilla, P.~Tett, P.A. Gillibrand, and M.~Inall.
\newblock Description and sensitivity analysis for the {L}{E}{S}{V} model:
  {W}ater quality variables and the balance of organisms in a fjordic region of
  restricted exchange.
\newblock {\em Ecological Modelling}, 220:2187--2205, 2009.

\bibitem{BurchardEtAl3}
H.~Burchard, K.~Bolding, W.~K\"{}uhn, A.~Meister, T.~Neumann, and L.~Umlauf.
\newblock Description of a flexible and extendable physical–biogeochemical
  model system for the water column.
\newblock {\em Journal of Marine Systems}, 61:180--211, 2006.

\bibitem{FennelNeumannBook1}
W.~Fennel and T.~Neumann.
\newblock {\em Introduction to the modelling of marine ecosystems}, volume~72
  of {\em Elsevier Oceanography Series}.
\newblock Elsevier, 2004.

\bibitem{Franks1}
P.J.S. Franks.
\newblock {N}{P}{Z} models of plankton dynamics: Their construction, coupling
  to physics, and application.
\newblock {\em Journal of Oceanography}, 58:379--387, 2002.

\bibitem{Anderson1}
T.R. Anderson1, A.P. Martin, R.S. Lampitt, C.N. Trueman, S.A. Henson, and D.J.
  Mayor.
\newblock Quantifying carbon fluxes from primary production to mesopelagic fish
  using a simple food web model.
\newblock {\em {I}{C}{E}{S} {J}ournal of Marine Science}, 76(3):690--701, 2019.

\bibitem{IbarraEtAl1}
D.A. Ibarra, K.~Fennel, and J.J. Cullen.
\newblock Coupling 3-{D} eulerian bio-physics ({R}{O}{M}{S}) with
  individual-based shellfish ecophysiology ({S}{H}{E}{L}{L}-{E}): {A} hybrid
  model for carrying capacity and environmental impacts of bivalve aquaculture.
\newblock {\em Ecological Modelling}, 273:63--78, 2014.

\bibitem{TettEtAl1}
P.~Tett, L.~Gilpin, H.~Svendsen, C.P. Erlandsson, U.~Larsson, S.~Kratzer,
  E.~Fouilland, C.~Janzen, J.Y. Lee, C.~Grenz, A.~Newton, J.~Gomes Ferreira,
  T.~Fernandes, and S.~Scory.
\newblock Eutrophication and some {E}uropean waters of restricted exchange.
\newblock {\em Continental Shelf Research}, 23(17-19):1635--1671, 2003.

\bibitem{ChenEtAl1}
W.B. Chen, W.C. Liu, and M.H. Hsu.
\newblock Water {Q}uality {M}odeling in a {T}idal {E}stuarine {S}ystem {U}sing
  a {T}hree-{D}imensional {M}odel.
\newblock {\em Environmental Engineering Science}, 28(6):443--459, 2011.

\bibitem{ReigadaEtAl1}
R.~Reigada, R.M. Hillary, M.A. Bees, J.M. Sancho, and F.~Sagu\'es.
\newblock Plankton blooms induced by turbulent flows.
\newblock {\em Proceedings of the Royal Society B}, 270:875--880, 2003.

\bibitem{BanasEtAl1}
N.S. Banas, J.~Zhang, R.G. Campbell, R.N. Sambrotto, M.W. Lomas, E.~Sherr,
  B.~Sherr, C.~Ashjian, D.~Stoecker, and E.J. Lessard.
\newblock Spring plankton dynamics in the {E}astern {B}ering {S}ea, 1971-2050:
  {M}echanisms of interannual variability diagnosed with a numerical model.
\newblock {\em Journal of Geophysical Research: Oceans}, 121(2):1476--1501,
  2016.

\bibitem{ChakrabortyFeudel1}
S.~Chakraborty and U.~Feudel.
\newblock Harmful algal blooms: combining excitability and competition.
\newblock {\em Theoretical Ecology}, 7:221--237, 2014.

\bibitem{LosEtAl1}
F.J. Los, M.T. Villars, and M.W.M.~Van der Tol.
\newblock A 3-dimensional primary production model ({B}{L}{O}{O}{M}/{G}{E}{M})
  and its applications to the (southern) {N}orth {S}ea (coupled
  physical-chemical-ecological model).
\newblock {\em Journal of Marine Systems}, 74(1-2):259--294, 2008.

\bibitem{HenseaBurchard1}
I.~Hensea and H.~Burchard.
\newblock Modelling cyanobacteria in shallow coastal seas.
\newblock {\em Ecological Modelling}, 221:238--244, 2010.

\bibitem{HenseBeckmann1}
I.~Hense and A.~Beckmann.
\newblock The representation of cyanobacteria life cycle processes in aquatic
  ecosystem models.
\newblock {\em Ecological Modelling}, 221:2330--2338, 2010.

\bibitem{HairerEtAlBook1}
E.~Hairer, C.~Lubich, and G.~Wanner.
\newblock {\em Geometric Numerical Integration. {S}tructure-Preserving
  Algorithms for Ordinary Differential Equations}.
\newblock Number~31 in Springer-Series in Computational Mathematics.
  Springer-Verlag, second edition, 2006.

\bibitem{HeinleSlawig2}
A.~Heinle and T.~Slawig.
\newblock Internal dynamics of {N}{P}{Z}{D} type ecosystem models.
\newblock {\em Ecological Modelling}, 254:33--42, 2013.

\bibitem{ZhangWang1}
T.~Zhang and W.~Wang.
\newblock {H}opf bifurcation and bistability of a
  nutrient-phytoplankton-zooplankton model.
\newblock {\em Applied Mathematical Modelling}, 36:6225--6235, 2012.

\bibitem{EdwardsBrindley1}
A.M.Edwards and J.~Brindley.
\newblock Zooplankton mortality and the dynamical behaviour of plankton
  population models.
\newblock {\em Bulletin of Mathematical Biology}, 61:303--339, 1999.

\bibitem{FennelEtAl1}
K.~Fennel, M.~Losch, J.~Schr\"{o}ter, and M.~Wenzel.
\newblock Testing a marine ecosystem model: sensitivity analysis and parameter
  optimization.
\newblock {\em Journal of Marine Systems}, 28:45--63, 2001.

\bibitem{BroekhuizenEtAl1}
N.~Broekhuizen, G.J. Rickard, J.~Bruggeman, and A.~Meister.
\newblock An improved and generalized second order, unconditionally positive,
  mass conserving integration scheme for biochemical systems.
\newblock {\em Applied Numerical Mathematics}, 58:319--340, 2008.

\bibitem{Edwards1}
A.M. Edwards.
\newblock Negative zooplankton do not exist--a response to `on the stability of
  some equilibrium points in a plankton population model'.
\newblock {\em Dynamical Systems}, 21(2):231--233, 2006.

\bibitem{KopeczMeister1}
S.~Kopecz and A.~Meister.
\newblock On order conditions for modified {P}atankar-{R}unge-{K}utta schemes.
\newblock {\em Applied Numerical Mathematics}, 123:159--179, 2018.

\bibitem{SchippmannBurchard1}
B.~Schippmann and H.~Burchard.
\newblock Rosenbrock methods in biogeochemical modelling - {A} comparison to
  {R}unge-{K}utta methods and modified {P}atankar schemes.
\newblock {\em Ocean Modelling}, 37:112--121, 2011.

\bibitem{BurchardEtAl4}
H.~Burchard, E.~Deleersnijder, and A.~Meister.
\newblock Application of modified {P}atankar schemes to stiff biogeochemical
  models for the water column.
\newblock {\em Ocean Dynamics}, 55:326--337, 2005.

\bibitem{Marsden2}
J.E. Marsden and W.~West.
\newblock Discrete mechanics and variational integrators.
\newblock {\em Acta Numerica}, 10:357--514, 2001.

\bibitem{DieleMarangi1}
F.~Diele and C.~Marangi.
\newblock Geometric numerical integration in ecological modelling.
\newblock {\em Mathematics}, 8(25):1--30, 2020.

\bibitem{RuckeltEtAl1}
J.~R\"{u}ckelt, V.~Sauerland, T.~Slawig, A.~Srivastav, B.~Ward, and
  C.~Patvardhan.
\newblock Parameter optimization and uncertainty analysis in a model of oceanic
  {C}{O}$_2$ uptake using a hybrid algorithm and algorithmic differentiation.
\newblock {\em Nonlinear Analysis: Real World Applications}, 11:3993--4009,
  2010.

\bibitem{Kriest1}
I.~Kriest.
\newblock Calibration of a simple and a complex model of global marine
  biogeochemistry.
\newblock {\em Biogeosciences}, 14:4965--4984, 2017.

\bibitem{CharbonneauKnappReport1}
P.~Charbonneau and B.~Knapp.
\newblock A {U}ser's {G}uide to {P}{I}{K}{A}{I}{A} 1.0, 1995.

\bibitem{HeinleSlawig3}
A.~Heinle and T.~Slawig.
\newblock Impact of parameter choice on the dynamics of {N}{P}{Z}{D} type
  ecosystem models.
\newblock {\em Ecological Modelling}, 267:93--101, 2013.

\bibitem{OschliesGarcon1}
A.~Oschlies and V.~Gar\c{c}on.
\newblock An eddy-permitting coupled physical-biological model of the {N}orth
  {A}tlantic. 1. {S}ensitivity to advection numerics and mixed layer physics.
\newblock {\em Global Biogeochemical Cycles}, 13(1):135--160, 1999.

\bibitem{HeinleSlawig1}
A.~Heinle and T.~Slawig.
\newblock {\em System Modeling and Optimization}, chapter Theoretical Analysis
  and Optimization of Nonlinear {O}{D}{E} Systems for Marine Ecosystem Models.
\newblock {C}{S}{M}{O} 2011. IFIP Advances in Information and Communication
  Technology. Springer, Berlin, Heidelberg, 2013.

\bibitem{SivanandamDeepaBook1}
S.N. Sivanandam and S.N. Deepa.
\newblock {\em Introduction to Genetic Algorithms}.
\newblock Springer-Verlag Berlin Heidelberg, 2008.

\bibitem{ArhonditsisEtAl1}
G.B. Arhonditsis, S.S. Qian, C.A. Stow, E.C. Lamon, and K.H. Reckhow.
\newblock Eutrophication risk assessment using bayesian calibration of
  process-based models: {A}pplication to a mesotrophic lake.
\newblock {\em Ecological Modelling}, 208:215--229, 2007.

\bibitem{ArhonditsisEtAl2}
G.B. Arhonditsis, D.~Papantou, W.~Zhang, G.~Perhar, E.~Massos, and M.~Shi.
\newblock Bayesian calibration of mechanistic aquatic biogeochemical models and
  benefits for environmental management.
\newblock {\em Journal of Marine Systems}, 73:8--30, 2008.

\bibitem{SchartauEtAl1}
M.~Schartau, P.~Wallhead, J.~Hemmings, U.~L\"{o}ptien, I.~Kriest, S.~Krishna,
  B.A. Ward, T.~Slawig, and A.~Oschlies.
\newblock Reviews and syntheses: parameter identification in marine planktonic
  ecosystem modelling.
\newblock {\em Biogeosciences}, 14:1647--1701, 2017.

\bibitem{FashamEtAl1}
M.J.R. Fasham, H.W. Ducklow, and S.M. McKelvie.
\newblock A nitrogen-based model of plankton dynamics in the oceanic mixed
  layer.
\newblock {\em Journal of Marine Research}, 48:591--639, 1990.

\bibitem{BurchardHofmeister1}
H.~Burchard and R.~Hofmeister.
\newblock A dynamic equation for the potential energy anomaly foranalysing
  mixing and stratification in estuaries and coastal seas.
\newblock {\em Estuarine, Coastal and Shelf Science}, 77:679--687, 2008.

\bibitem{WardDutkiewiczFollows1}
B.A. Ward, S.~Dutkiewicz, and M.J. Follows.
\newblock Modelling spatial and temporal patterns in size-structured marine
  plankton communities: top–down and bottom–up controls.
\newblock {\em Journal of Plankton Research}, 36(1):31--47, 2014.

\bibitem{OschliesKoeveGarcon2}
A.~Oschlies, W.~Koeve, and V.~Gar\c{c}on.
\newblock An eddy-permitting coupled physical-biological model of the {N}orth
  {A}tlantic. 2. {E}cosystem dynamics and comparison with satellite and
  {J}{G}{O}{F}{S} local studies data.
\newblock {\em Global Biogeochemical Cycles}, 14(1):499--523, 2000.

\bibitem{Strang1}
G.~Strang.
\newblock Approximating semigroups and the consistency of difference schemes.
\newblock 20(1):1--7, 1969.

\bibitem{BodeEtAl1}
A.~Bode, J.A. Botas, and E.~Fern\'andez.
\newblock Nitrate storage by phytoplankton in a coastal upwelling environment.
\newblock {\em Marine Biology}, 129:399--406, 1997.

\bibitem{EscribanoRodriguez1}
R.~Escribano and L.~Rodr\'iguez.
\newblock Seasonal size variation and growth of {C}alanus chilensis {B}rodsky
  in northern {C}hile.
\newblock 68:373--382, 1995.

\bibitem{Gismervik1}
I.~Gismervik.
\newblock Stoichiometry of some marine planktonic crustaceans.
\newblock {\em Journal of Plankton Research}, 19(2):279--285, 1997.

\bibitem{Thornton1}
D.C.O. Thornton.
\newblock Dissolved organic matter ({D}{O}{M}) release by phytoplankton in the
  contemporary and future ocean.
\newblock {\em European Journal of Phycology}, 49(1):20--46, 2014.

\bibitem{HarveyKratzerAndersson1}
E.T. Harvey, S.~Kratzer, and A.~Andersson.
\newblock Relationships between colored dissolved organic matter and dissolved
  organic carbon in different coastal gradients of the {B}altic {S}ea.
\newblock {\em AMBIO}, 44:392--401, 2015.

\bibitem{MarkagerEtAl1}
S.~Markager, C.A. Stedmon, and M.~Sondergaard.
\newblock Seasonal dynamics and conservative mixing of dissolved organic matter
  in the temperate eutrophic estuary {H}orsens fjord.
\newblock {\em Estuarine, Coastal and Shelf Science}, 92:376--388, 2011.

\end{thebibliography}
\end{document}